\RequirePackage{lineno}
\documentclass[a4paper, 11pt, epsfig]{article}

\usepackage{jheppub} 

\usepackage[T1]{fontenc} 
\usepackage{graphicx,color,dcolumn,booktabs,bm}
\usepackage{longtable,lscape}
\usepackage{amsmath}
\usepackage{indentfirst}
\usepackage{epsfig}
\usepackage{epstopdf} 
\usepackage{slashed}  
\usepackage{color}
\usepackage{multirow}
\usepackage{soul}     
\usepackage{graphicx,color,dcolumn,booktabs,bm}
\usepackage{verbatim}
\usepackage{orcidlink}

\usepackage{multirow}
\usepackage{makecell}
\usepackage{array}
\usepackage{color}

\newcommand{\lum}{{\cal L}}
\newcommand{\eff}{\varepsilon}
\newcommand{\BR}{{\cal B}}
\newcommand{\pip}{\pi^+}
\newcommand{\pim}{\pi^-}

\newcommand{\EE}{e^+e^-}

\newcommand{\LL}{\ell^+\ell^-}

\renewcommand\tabcolsep{2.0pt}

\title{\boldmath Observations of the singly Cabibbo-suppressed decays $\Xi_c^{+} \to pK_{S}^{0}$, $\Xi_c^+ \to \Lambda \pi^+$,
and $\Xi_c^+ \to \Sigma^{0} \pi^+$ at Belle and Belle II}
	
\preprint{\vbox{\hbox{   }
		\hbox{}
		\hbox{}
		\hbox{Belle II Preprint 2024-028}
		\hbox{KEK Preprint 2024-29}}}

\collaboration{The Belle and Belle II Collaborations}
\author{I.~Adachi\,\orcidlink{0000-0003-2287-0173},} 
\author{L.~Aggarwal\,\orcidlink{0000-0002-0909-7537},} 
\author{N.~Akopov\,\orcidlink{0000-0002-4425-2096},} 
\author{M.~Alhakami\,\orcidlink{0000-0002-2234-8628},} 
\author{A.~Aloisio\,\orcidlink{0000-0002-3883-6693},} 
\author{N.~Althubiti\,\orcidlink{0000-0003-1513-0409},} 
\author{N.~Anh~Ky\,\orcidlink{0000-0003-0471-197X},} 
\author{D.~M.~Asner\,\orcidlink{0000-0002-1586-5790},} 
\author{H.~Atmacan\,\orcidlink{0000-0003-2435-501X},} 
\author{T.~Aushev\,\orcidlink{0000-0002-6347-7055},} 
% ...

\abstract{Using data samples of 983.0~$\rm fb^{-1}$ and 427.9~$\rm fb^{-1}$ accumulated with the Belle 
and Belle~II detectors operating at the KEKB and SuperKEKB asymmetric-energy $\EE$ colliders, singly 
Cabibbo-suppressed decays $\Xi_c^{+} \to pK_{S}^{0}$, $\Xi_c^+ \to \Lambda \pi^+$, and $\Xi_c^+ \to \Sigma^{0} \pi^+$ 
are observed for the first time. The ratios of branching 
fractions of $\Xi_{c}^{+}\to p K_{S}^{0}$, $\Xi_{c}^{+}\to \Lambda \pi^{+}$, and $\Xi_{c}^{+}\to \Sigma^{0} \pi^{+}$
relative to that of $\Xi_c^+ \to \Xi^- \pi^{+} \pi^{+}$ are measured to be
\begin{equation}
	\frac{\BR(\Xi_c^+ \to pK_S^0)}{\BR(\Xi_c^{+} \to \Xi^{-} \pip \pip)} = (2.47 \pm 0.16 \pm 0.07)\% \notag,
\end{equation}
\begin{equation}
	\frac{\BR(\Xi_c^+ \to \Lambda \pi^+)}{\BR(\Xi_c^{+} \to \Xi^{-} \pip \pip)} = (1.56 \pm 0.14 \pm 0.09)\% \notag,
\end{equation}
\begin{equation}
	\frac{\BR(\Xi_c^+ \to \Sigma^0 \pi^+)}{\BR(\Xi_c^{+} \to \Xi^{-} \pip \pip)} = (4.13 \pm 0.26 \pm 0.22)\% \notag.
\end{equation} 
Multiplying these values by the branching fraction of the normalization channel, $\BR(\Xi_c^{+} \to \Xi^{-} \pip\pip) = (2.9 \pm 1.3)\%$,
the absolute branching fractions are determined to be 
\begin{equation}
	\BR(\Xi_c^{+} \to p K_{S}^{0}) = (7.16 \pm 0.46 \pm 0.20 \pm 3.21) \times 10^{-4} \notag,
\end{equation}
\begin{equation}
	\BR(\Xi_c^{+} \to \Lambda \pip) = (4.52 \pm 0.41 \pm 0.26 \pm 2.03) \times 10^{-4} \notag,
\end{equation}
\begin{equation}
	\BR(\Xi_c^{+} \to \Sigma^0 \pip) = (1.20 \pm 0.08 \pm 0.07 \pm 0.54) \times 10^{-3} \notag.
\end{equation}
The first and second uncertainties above are statistical and systematic, respectively,
while the third ones arise from the uncertainty in $\BR(\Xi_c^{+} \to \Xi^{-} \pi^{+} \pi^{+})$.}


\begin{document} 
\begin{sloppypar}
\maketitle
\flushbottom

\section{Introduction}
\noindent The study of charmed baryons is valuable for exploring the subtle interplay between the strong and weak 
interactions. In hadronic weak decays of charmed baryons, nonfactorizable contributions from the internal $W$-emission
and $W$-exchange diagrams play an essential role and cannot be neglected. This is unlike the 
situation in heavy meson decay where they are negligible~\cite{Cheng:2021qpd}. In particular, there exist decay
channels that receive only nonfactorizable contributions, such as $\Lambda_c^+ \to \Xi^0 K^+$ and $\Xi_c^0 \to \Sigma^+ K^-$. 
Therefore, studying nonfactorizable effects is critical for understanding the dynamics of charmed baryon decays.

\begin{figure}[htbp]
	\begin{center}
		\begin{minipage}{0.32\textwidth}
		\centerline{\includegraphics[width=5cm]{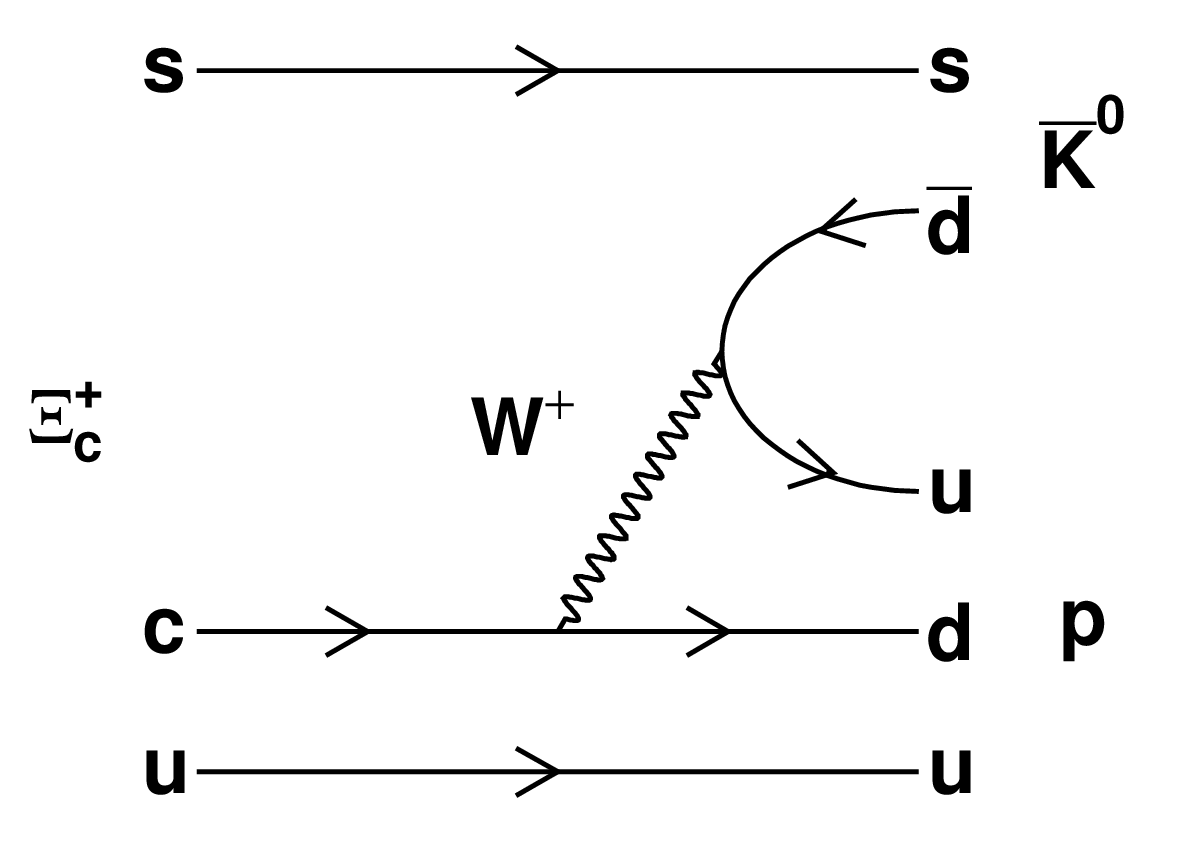}}
		\end{minipage}
		\begin{minipage}{0.32\textwidth}
		\centerline{\includegraphics[width=5cm]{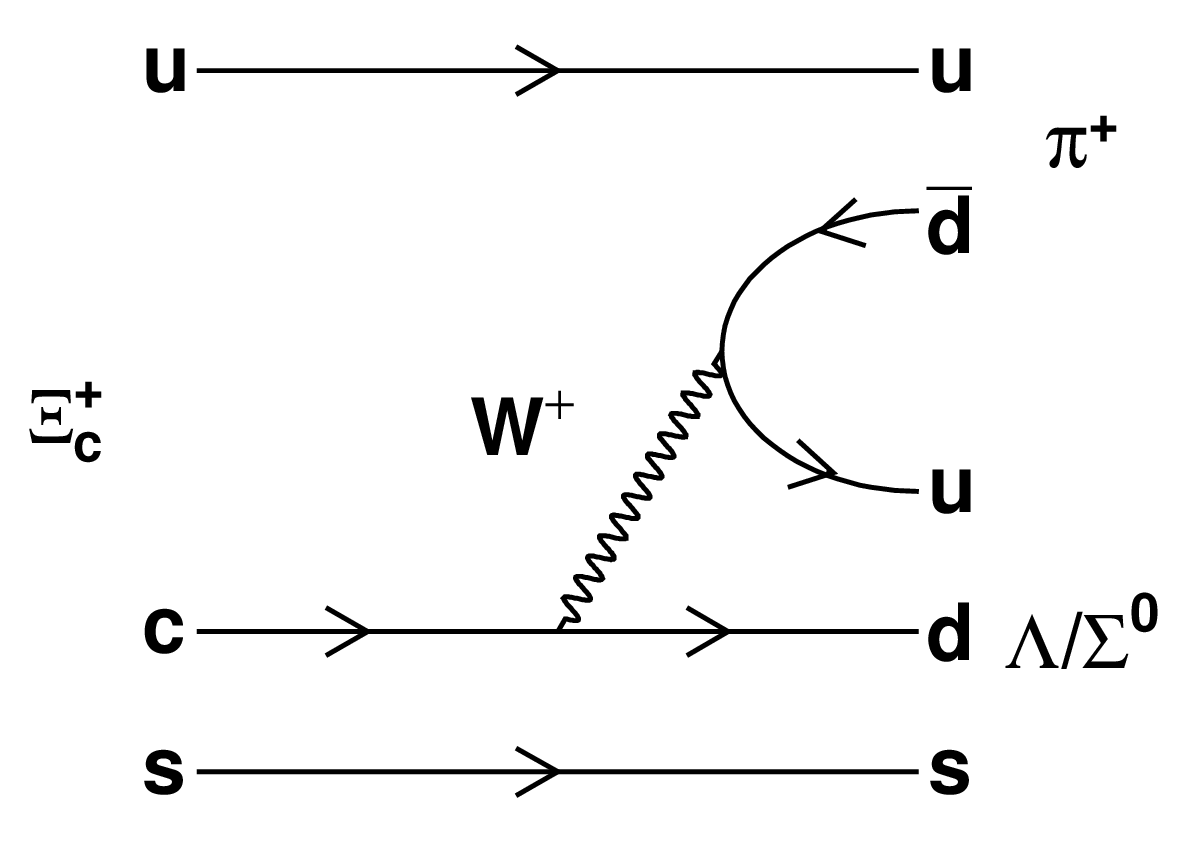}}
		\end{minipage}
		\begin{minipage}{0.32\textwidth}
     	\centerline{\includegraphics[width=5cm]{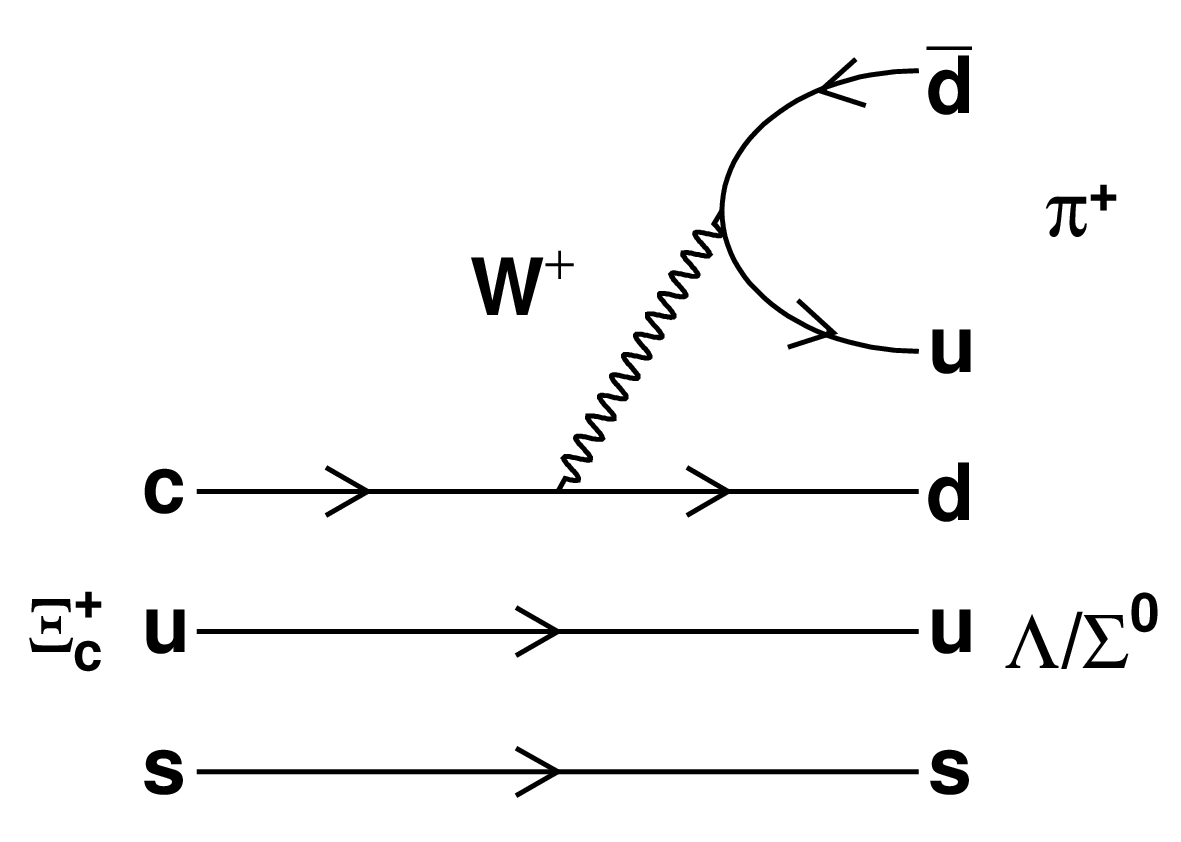}}
        \end{minipage}
    	\put(-400,55){\bf (a)} \put(-260,55){\bf (b)} \put(-120,55){\bf (c)}
    
        \vspace{0.5cm}
    
        \begin{minipage}{0.32\textwidth}
     	\centerline{\includegraphics[width=5cm]{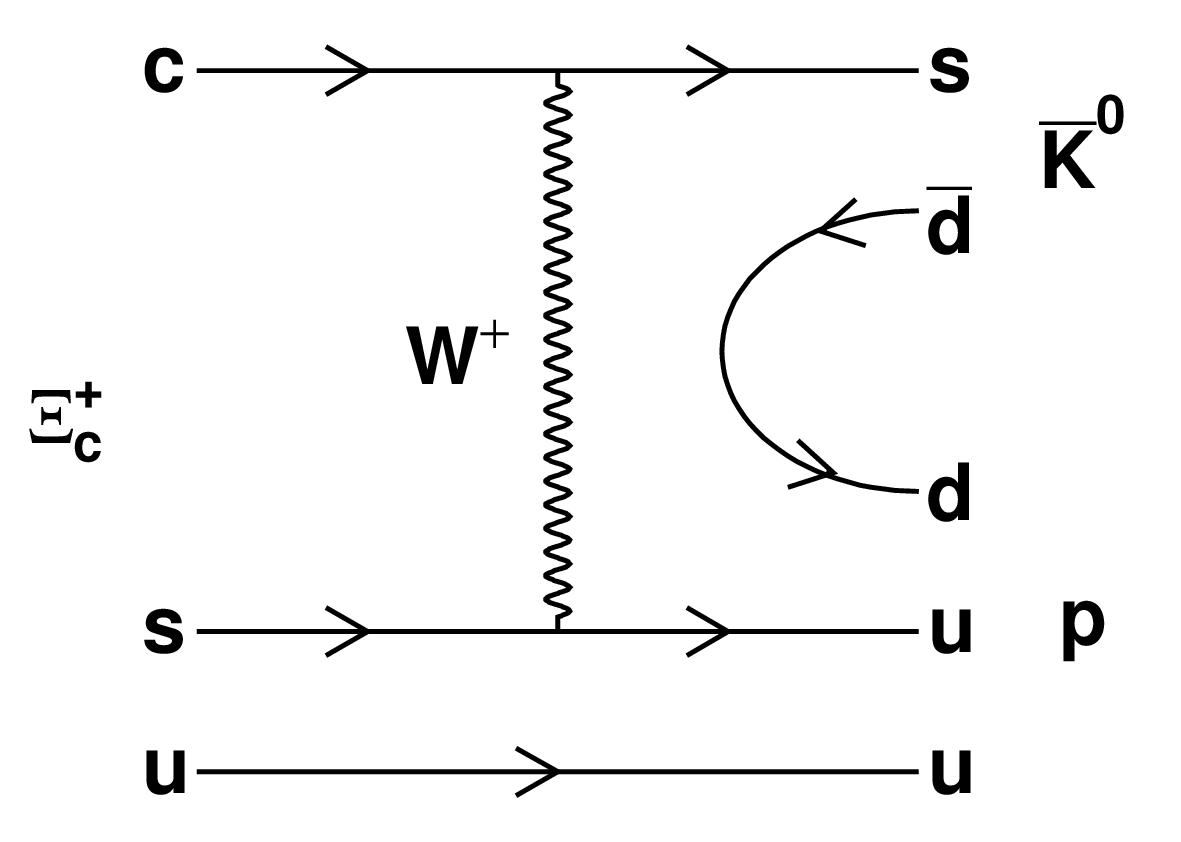}}
        \end{minipage}
        \begin{minipage}{0.32\textwidth}
        \centerline{\includegraphics[width=5cm]{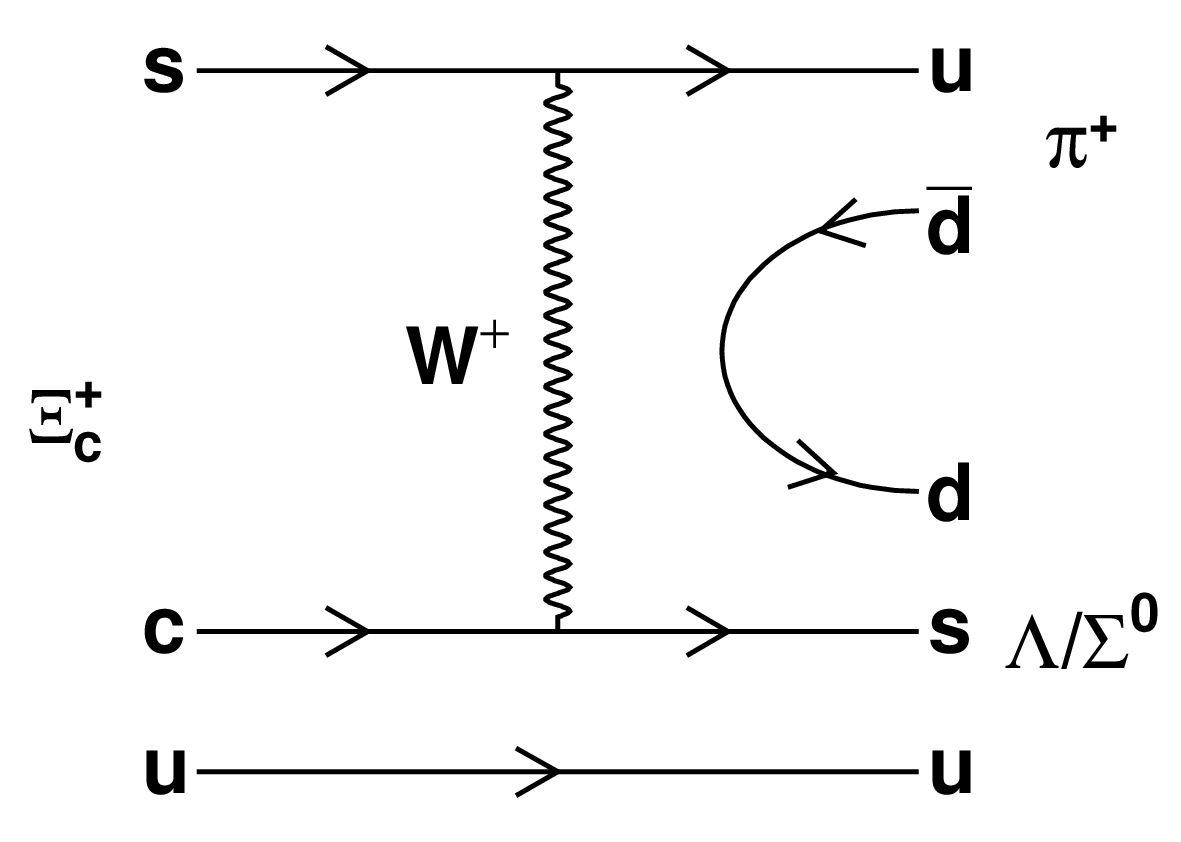}}
        \end{minipage}
		\put(-260,55){\bf (d)} \put(-120,55){\bf (e)} 
		\caption{Representative decay diagrams for (a,b) internal $W$-emission, (c) external $W$-emission, and (d,e) $W$-exchange contributions for the singly Cabibbo-suppressed decays $\Xi_c^+ \to p\bar{K}^{0}$, $\Xi_c^+ \to \Lambda \pi^+$, and $\Xi_c^+ \to \Sigma^0 \pi^+$.}\label{Fig1}
	\end{center}
\end{figure}

In the last few years, there has been a significant advance in the experimental and theoretical studies of 
hadronic weak decays of anti-triplet charmed baryons ($\Lambda_c^+$, $\Xi_c^0$, and $\Xi_c^+$)~\cite{Cheng:2021qpd, ParticleDataGroup:2024cfk}. Notably, the absolute
branching fractions of $\Xi_c^{0} \to \Xi^{-} \pip$ and $\Xi_c^{+} \to \Xi^{-} \pip\pip$ have been measured by Belle to be 
$\BR(\Xi_c^{0} \to \Xi^{-} \pip) =  (1.80 \pm 0.50 (\rm stat.) \pm 0.14 (\rm syst.))\%$~\cite{Belle:2018kzz} and
$\BR(\Xi_c^{+} \to \Xi^{-} \pip\pip) = (2.86 \pm 1.21 (\rm stat.) \pm 0.38 (\rm syst.))\%$~\cite{Belle:2019bgi}.
Most measurements of $\Xi_c^0$ and $\Xi_c^+$ branching fractions are measured relative to these two decay modes. 
The measurements of these absolute branching fractions have 
sparked renewed interest in the study of $\Xi_c^0$ and $\Xi_c^+$ decays~\cite{Belle:2020ito, Belle:2021crz, Belle:2021zsy, Belle:2021avh,Belle:2022raw,Belle:2023ngs,Belle-II:2024jql}. Comprehensive and precise experimental measurements are 
essential to test different theoretical models and illuminate the decay mechanisms of anti-triplet charmed baryons.
Theoretical calculations for the two-body hadronic weak decays of $\Xi_c^+$ 
have been performed based on dynamical model calculations~\cite{Zou:2019kzq}
and $\rm SU(3)_F$ flavor symmetry methods~\cite{Zhao:2018mov,Geng:2018plk,Geng:2019xbo,Huang:2021aqu,Hsiao:2021nsc,Zhong:2022exp,Xing:2023dni,Geng:2023pkr,Liu:2023dvg,Zhong:2024qqs}.
However, most of these decay channels have not yet been measured experimentally, especially the singly Cabibbo-suppressed decay modes. 
Figure~\ref{Fig1} shows the typical decay diagrams for the singly Cabibbo-suppressed decays $\Xi_c^+ \to p\bar{K}^{0}$,
$\Xi_c^+ \to \Lambda \pi^+$, and $\Xi_c^+ \to \Sigma^0 \pi^+$.
Note that the decay $\Xi_c^+ \to p\bar{K}^{0}$ cannot proceed via an external $W$-emission diagram and so 
it occurs solely through nonfactorizable diagrams. Thus, measuring this decay enables a direct evaluation of 
the significance of nonfactorizable contributions in $\Xi_c^+$ decay.
The branching fractions of these three decay channels are predicted by different theoretical models to cover the range of $10^{-4}$
to $10^{-3}$~\cite{Zou:2019kzq, Zhao:2018mov,Geng:2018plk,Geng:2019xbo,Huang:2021aqu,Hsiao:2021nsc,Zhong:2022exp,Xing:2023dni,Geng:2023pkr,Liu:2023dvg,Zhong:2024qqs}.

In this paper, we study the singly Cabibbo-suppressed decays $\Xi_c^{+} \to pK_{S}^{0}$, $\Xi_c^+ \to \Lambda \pi^+$,
and $\Xi_c^+ \to \Sigma^{0} \pi^+$ for the first time.
Using $\Xi_c^+ \to \Xi^- \pi^+ \pi^+$ as the normalization channel, we measure the ratios of branching fractions of 
$\Xi_{c}^{+}\to p K_{S}^{0}$, $\Xi_{c}^{+}\to \Lambda \pi^{+}$, 
and $\Xi_{c}^{+}\to \Sigma^{0} \pi^{+}$ relative to that of $\Xi_c^+ \to \Xi^- \pi^{+} \pi^{+}$.
This analysis is based on data samples of 983.0~$\rm fb^{-1}$~\cite{Belle2} and 427.9~$\rm fb^{-1}$~\cite{Belle-II:2024vuc}
accumulated with the Belle and Belle~II detectors operating at the KEKB and SuperKEKB 
asymmetric-energy $\EE$ colliders, respectively. Charge-conjugate modes are included throughout the paper.

\section{The Belle and Belle~II detectors and data samples}
\noindent The Belle detector~\cite{Belle1}, which operated between 1999 and 2010 at the KEKB asymmetric-energy
$e^+e^-$ collider~\cite{KEKB1,KEKB2}, was a large cylindrical solid-angle magnetic spectrometer that 
consisted of a silicon vertex detector, a 50-layer central drift chamber (CDC), an array of aerogel threshold Cherenkov
counters (ACC), time-of-flight scintillation counters (TOF), and an electromagnetic calorimeter composed of CsI(Tl) 
crystals located inside a superconducting solenoid coil that provides a $1.5~\hbox{T}$ magnetic field. 
An iron flux return equipped with resistive plate chambers located outside the coil is instrumented to detect
$K^{0}_{L}$ mesons and to identify muons. The detector is described in detail elsewhere~\cite{Belle1}.

The Belle II detector~\cite{BelleII} is located at the interaction point (IP) of the SuperKEKB asymmetric-energy
$\EE$ collider~\cite{SKEKB}, and has been collecting data since 2019.
The Belle~II detector is based on the Belle detector but contains
several new subsystems, as well as substantial upgrades to others. The innermost
subdetector is the vertex detector (VXD) which includes two inner layers of pixel
sensors and four outer layers of double-sided silicon microstrip sensors. Charged particle momenta and charges are measured by 
a new large-radius, helium-ethane, small-cell CDC, which also offers charged-particle-identification information
through a measurement of specific ionization. A Cherenkov-light angle and time-of-propagation (TOP) detector 
surrounding the CDC provides charged-particle identification in the central detector volume, 
supplemented by proximity-focusing, aerogel, ring-imaging Cherenkov (ARICH) detectors in the 
forward region with respect to the electron beam. The Belle CsI(Tl) crystal electromagnetic
calorimeter, the Belle solenoid and iron flux return are reused in the Belle II detector.
The electromagnetic calorimeter readout electronics have been upgraded and the instrumentation
in the flux return to identify $K_L^{0}$ mesons and muons has been replaced.

This measurement uses data recorded at center-of-mass (c.m.) energies at or near the
$\Upsilon(nS)~ (n=1,2,3,4,5)$ resonances by the Belle detector, and at or near
the $\Upsilon(4S)$ and at $\sqrt{s}=10.75$~GeV by the Belle II detector. The data samples correspond to
integrated luminosities of 983.0~$\rm fb^{-1}$~\cite{Belle2} and 427.9~$\rm fb^{-1}$~\cite{Belle-II:2024vuc} 
with Belle and Belle II, respectively.

Monte Carlo (MC) signal events are generated using {\textsc EvtGen}~\cite{EVTGEN} 
and used to optimize signal selection criteria and calculate the reconstruction efficiencies. 
Continuum $e^+ e^- \to c\bar{c}$ events are generated using {\sc PYTHIA6}~\cite{PYTHIA1} for Belle
and {\textsc PYTHIA8}~\cite{PYTHIA2} and {\textsc KKMC}~\cite{kkmc} for Belle~II, where one of the two charm quarks 
hadronizes into a $\Xi_c^+$ baryon. The decays $\Xi_c^+ \to pK_S^0/\Lambda\pi^+/\Sigma^0\pi^+/\Xi^-\pi^+\pi^+$
are generated using a phase space model. The effect of final-state radiation is taken into
account in the simulation using the {\textsc PHOTOS} package~\cite{PHOTOS}. The simulated signal
events are processed with detector simulations based on {\sc GEANT3}~\cite{GEANT3} for the Belle 
detector and {\sc GEANT4}~\cite{GEANT4} for the \mbox{Belle II} detector.

Inclusive MC samples of $\Upsilon(1S,~2S,~3S)$ decays, $\Upsilon(4S) \to B\bar{B}$,
$\Upsilon(5S) \to B_{(s)}^{(*)} \bar{B}_{(s)}^{(*)}$, and $e^+e^- \to q\bar{q}$ ($q=u,\,d,\,s,\,c$)
at c.m.~energies of 10.520, 10.580, and 10.867~GeV are used to optimize signal selection criteria
and study the composition of backgrounds in the Belle analysis, 
corresponding to twice the integrated luminosity of the Belle data.
In the Belle II analysis, we use inclusive MC samples of $e^+e^- \to q\bar{q}$ 
at c.m.~energies of 10.520, 10.580 and 10.750~GeV and $\Upsilon(4S) \to B\bar{B}$ corresponding to
four times the integrated luminosity of the Belle~II data, to optimize signal selection criteria
and study the backgrounds~\cite{topo}.

\section{Event selection criteria}
\noindent We reconstruct the decay modes $\Xi_c^+ \to pK_S^0$, $\Lambda \pi^+$, $\Sigma^0 \pi^+$, and
$\Xi^- \pi^+ \pi^+$, followed by the decays $K_S^0 \to \pi^+ \pi^-$, $\Sigma^0 \to \Lambda \gamma$, 
$\Xi^- \to \Lambda \pi^-$, and $\Lambda \to p \pi^-$. We use the Belle II analysis 
software framework (BASF2) to reconstruct the events at Belle and Belle II~\cite{basf2}. The Belle II data
are directly processed with this framework, 
while the tracks and clusters in the processed Belle data are converted to BASF2 format using the B2BII 
software package~\cite{b2bii}. After conversion, the same reconstruction software is applied to both data samples.
The event selection criteria described below are optimized by maximizing the 
figure of merit $N_{\rm sig}/\sqrt{N_{\rm sig} + N_{\rm bkg}}$. 
Here, $N_{\rm sig}$ represents the number of expected $\Xi_c^+ \to pK_S^0/\Lambda \pi^+/\Sigma^0 \pi^+$ signal events, based on the
branching fraction predicted in ref.~\cite{Zou:2019kzq}, and $N_{\rm bkg}$ denotes the number of 
background events in the $\Xi_c^+$ signal region, obtained from the inclusive MC samples and scaled by
the ratio of yields between data and inclusive MC in the normalized $\Xi_c^+$ sideband regions.
The optimal selection criteria are not significantly dependent on the choice of theoretically predicted branching fractions.
The signal region for the $\Xi_c^+$ is defined as
$|M(pK_S^0/\Lambda \pi^+/\Sigma^0 \pi^+) - m_{\Xi_c^+}| < 20~{\rm MeV}/c^2$ 
(approximately 3 standard deviations, $\sigma$),
and the sideband regions are defined as $32~{\rm MeV}/c^2 < |M(pK_S^0/\Lambda \pi^+/\Sigma^0 \pi^+) - m_{\Xi_c^+}| < 52~{\rm MeV}/c^2$. 
Here and throughout this paper, $m_{i}$ represents the known mass of the particle $i$~\cite{ParticleDataGroup:2024cfk}.
We apply nearly identical event selection criteria in the Belle and Belle II analyses unless otherwise stated.

The impact parameters of charged tracks, except for those of the decay products of $K_S^0$, 
$\Lambda$, and $\Xi^-$, measured with respect to the $\EE$ IP, are required to be less than 0.2~cm
perpendicular to the $z$-axis and less than 1~cm parallel to it. The $z$-axis is defined as the central solenoid axis
with the positive direction toward the $e^-$ beam, common to both the Belle and Belle~II detectors.
For the particle identification (PID) of a charged track, information from different detector subsystems, 
including specific ionization in the CDC, time measurement in the TOF (TOP), and the response of the ACC (ARICH)
of Belle (Belle II), is combined to form a likelihood ratio,  
$\mathcal{R}(h|h^{\prime}) = \mathcal{L}(h)/[\mathcal{L}(h) + \mathcal{L}(h^{\prime})]$,
where $\mathcal{L}(h^{(\prime)})$ is the likelihood of the charged track 
being a hadron $h^{(\prime)} =$ $p$, $K$, or $\pi$ as appropriate. 
Tracks with $\mathcal{R}(p|K) > 0.6$ and
$\mathcal{R}(p|\pi) > 0.6$ are identified as proton candidates; charged pion candidates must satisfy $\mathcal{R}(\pi|K) > 0.6$ 
with an average efficiency of 91\% (90\%) in Belle (Belle~II), while 6\% (7\%) of kaons are misidentified as pions.
To suppress backgrounds from low-momentum protons and pions, we require the momentum of the proton in the laboratory 
frame to be greater than 1.1~GeV/$c$ for the $\Xi_c^+ \to p K_S^0$ mode and the pion momentum to be greater than
0.6~GeV/$c$ and 0.4~GeV/$c$ for the $\Xi_c^+ \to \Lambda \pi^+$ and $\Xi_c^+ \to \Sigma^0 \pi^+$ modes, respectively.

The $K_{S}^{0}$ candidates are first reconstructed from pairs of oppositely charged tracks, 
which are treated as pions. In the Belle analysis, we use an artificial neural network~\cite{Feindt:2006pm}
based on two sets of input variables~\cite{input} to select the $K_{S}^{0}$ candidates. In the
Belle~II analysis, the significance of flight distance ($L_{\rm fl}/\sigma_{L_{\rm fl}}$) of $K_S^0$ is required to be greater than 10
to suppress combinatorial backgrounds.
Here and below, the flight distance ($L_{\rm fl}$) of a particle is calculated as the projection of the displacement vector, 
which joins its production and decay vertices, onto the direction of its momentum. The corresponding uncertainty 
($\sigma_{L_{\rm fl}}$) is calculated by propagating the uncertainties in the vertex positions and momenta.
The signal region of the reconstructed $K_S^0$ candidates is defined as $|M(\pi^+\pi^-) - m_{K_S^0}| < 10$~MeV/$c^2$ ($\sim$3$\sigma$).

The $\Lambda$ candidates are reconstructed via the decay $\Lambda \to p \pi^-$.
In the Belle analysis, we select $\Lambda$ candidates using $\Lambda$-momentum-dependent criteria 
based on four parameters: the distance between the two daughter tracks along the $z$-axis at their closest approach; 
the minimum distance between the daughter tracks and the IP in the transverse plane; the angular difference 
between the $\Lambda$ flight direction and the direction between the IP and the $\Lambda$ decay vertex in
the transverse plane; and the flight length of the $\Lambda$ in the transverse plane. In the Belle II analysis, 
the significance of the flight distance of the $\Lambda$ is required to be $L_{\rm fl}/\sigma_{L_{\rm fl}} > 10$ to suppress 
combinatorial backgrounds. The signal region of the reconstructed $\Lambda$ candidates is defined
as $|M(p\pi^-) - m_{\Lambda}| <3.5$~MeV/$c^2$ ($\sim$3$\sigma$).

An ECL cluster is used as a photon candidate if it is not consistent with the extrapolated path
of any charged track.
To suppress background from neutral hadrons, we require $E(3\times3)$/$E(5\times5)$ $\geq$ 85\% 
where $E(n\times n)$ is the energy contained in an $n\times n$ crystal region centered on the crystal with the highest energy
(for Belle II only, the outer corner crystals are not included).
The photon energy must exceed 80~MeV in the laboratory frame to further suppress the combinatorial
backgrounds. The selected photon candidate is then combined with a $\Lambda$ candidate to form a $\Sigma^0$ candidate. 
The signal region of the reconstructed  $\Sigma^0$ candidates is defined
as $|M(\Lambda \gamma) - m_{\Sigma^0}| < 6$~MeV/$c^2$ ($\sim$2$\sigma$).

In the reconstruction of $\Xi^- \to \Lambda \pi^-$, the selected $\Lambda$ candidate is combined 
with a $\pi^-$ to form a $\Xi^-$ candidate. The $\pi^-$ is not required to satisfy any PID criteria
as the expected kinematics of the $\Xi^-$ signal gives sufficient discrimination,
but its transverse momentum must exceed 50 MeV/$c$ to eliminate background from low-momentum pions.
Additionally, the distance from the IP to the $\Xi^-$ decay vertex must be less than that to the
$\Lambda$ decay vertex. A vertex fit is applied to the entire $\Xi^-$ decay chain~\cite{Belle-IIanalysissoftwareGroup:2019dlq},
including subsequent decay products, with the $p \pi^-$ invariant
mass constrained to the known $\Lambda$ mass~\cite{ParticleDataGroup:2024cfk}. 
The signal region of the reconstructed  $\Xi^-$ candidates is defined
as $|M(\Lambda \pi^-) - m_{\Xi^-}| < 6$~MeV/$c^2$ ($\sim$3$\sigma$).

The $pK_S^0$, $\Lambda \pi^+$, $\Sigma^0 \pi^+$, and $\Xi^- \pi^+ \pi^+$ combinations are used to form 
$\Xi_c^+$ candidates. A vertex-fitting algorithm is applied to the entire decay chain, incorporating
mass constraints for the intermediate states and ensuring that the $\Xi_c^+$ originates from the IP~\cite{Belle-IIanalysissoftwareGroup:2019dlq}.
The goodness-of-fit $\chi^2$ is required to be less than 20 for the $\Xi_c^+ \to pK_S^0/\Lambda\pi^+/\Sigma^0\pi^+$ 
modes and less than 100 for the $\Xi_c^+ \to \Xi^- \pi^+ \pi^+$ mode. 
The significance of the $\Xi_c^+$ flight distance is required to be greater than 1.5 (3.0) for the Belle (Belle~II) analysis.
This criterion suppresses a significant number of background events,
particularly in the Belle II data which benefits from the superior vertex
resolution of the Belle~II VXD detector and the smaller beam spot of SuperKEKB.

To reduce combinatorial backgrounds, especially from $B$-meson decays, the scaled momentum
$x_{p} = p^{*}_{\Xi_{c}^+}$/$p_{\rm max}$ is required to be larger than 0.55 in both
Belle and Belle~II analyses. Here, $p^{*}_{\Xi_{c}^+}$ is the momentum of $\Xi_{c}^+$
candidates in the $\EE$ c.m.\ frame, and $p_{\rm max}=\frac{1}{c}\sqrt{E^2_{\rm beam}-M_{\Xi_c^+}^2 c^4}$,
where $E_{\rm beam}$ is the beam energy in the $\EE$ c.m.\ frame and $M_{\Xi_{c}^+}$
is the invariant mass of $\Xi_{c}^+$ candidates. Finally, if there are multiple $\Xi_c^+$ candidates
in an event, all the combinations are retained for further analysis.
The fractions of events that have multiple candidate events in signal MC simulations for Belle (Belle II)
are 0.2\% (0.3\%), 0.4\% (0.7\%), 2.7\% (2.8\%), and 2.0\% (1.6\%) for the
$\Xi_c^+ \to pK_S^0$, $\Xi_c^+ \to \Lambda \pi^+$, $\Xi_c^+ \to \Sigma^0 \pi^+$, and 
$\Xi_c^+ \to \Xi^- \pi^+ \pi^+$ decay modes, respectively. These values are consistent with the multiple
candidate rates observed in the data.

\section{Branching fractions of $\Xi_c^+ \to pK_S^0$, $\Xi_c^+ \to \Lambda \pi^+$, and $\Xi_c^+ \to \Sigma^0 \pi^+$ decays}
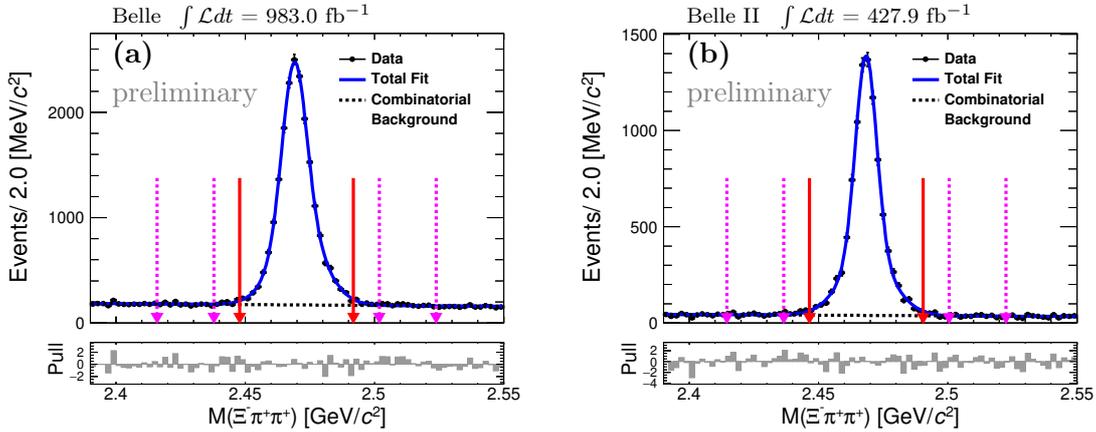
\begin{figure}[htbp]
	\begin{center}
		\begin{minipage}{0.49\textwidth}
			\centerline{\includegraphics[width=8cm]{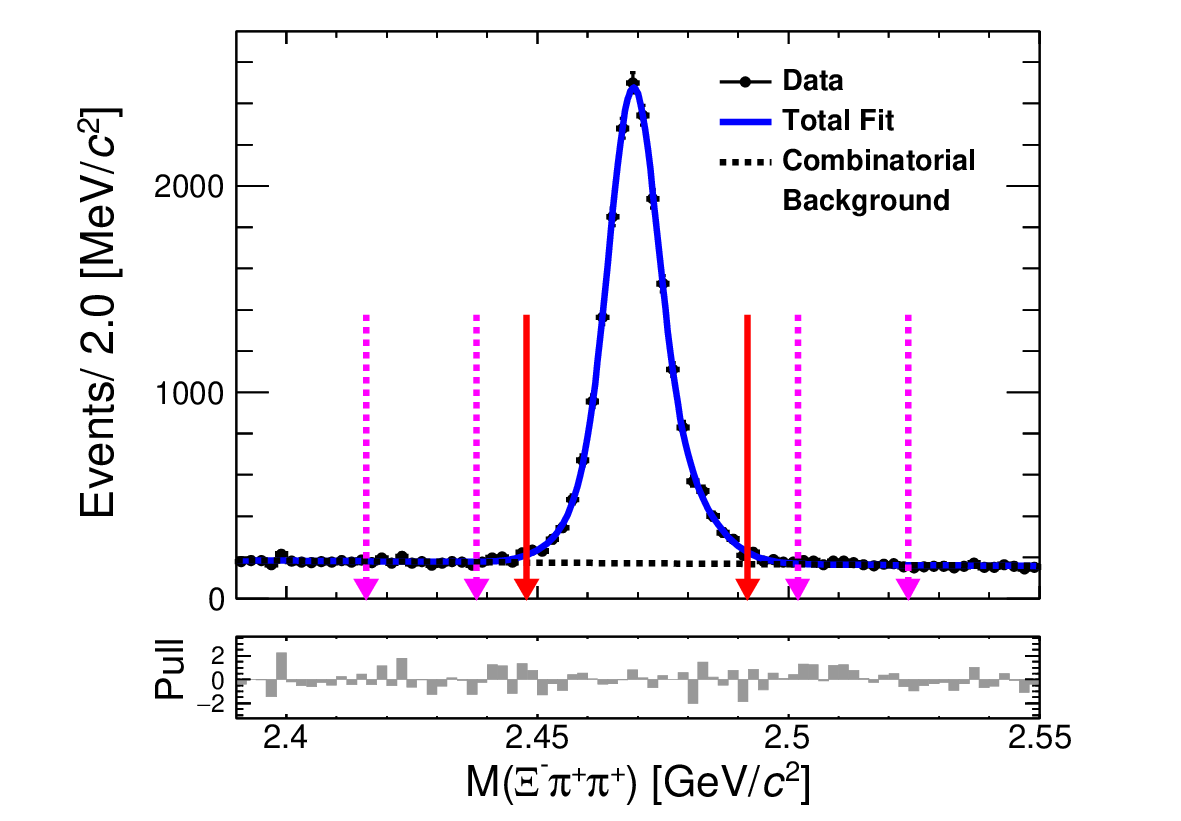}} 
		\end{minipage}
		\begin{minipage}{0.49\textwidth}
			\centerline{\includegraphics[width=8cm]{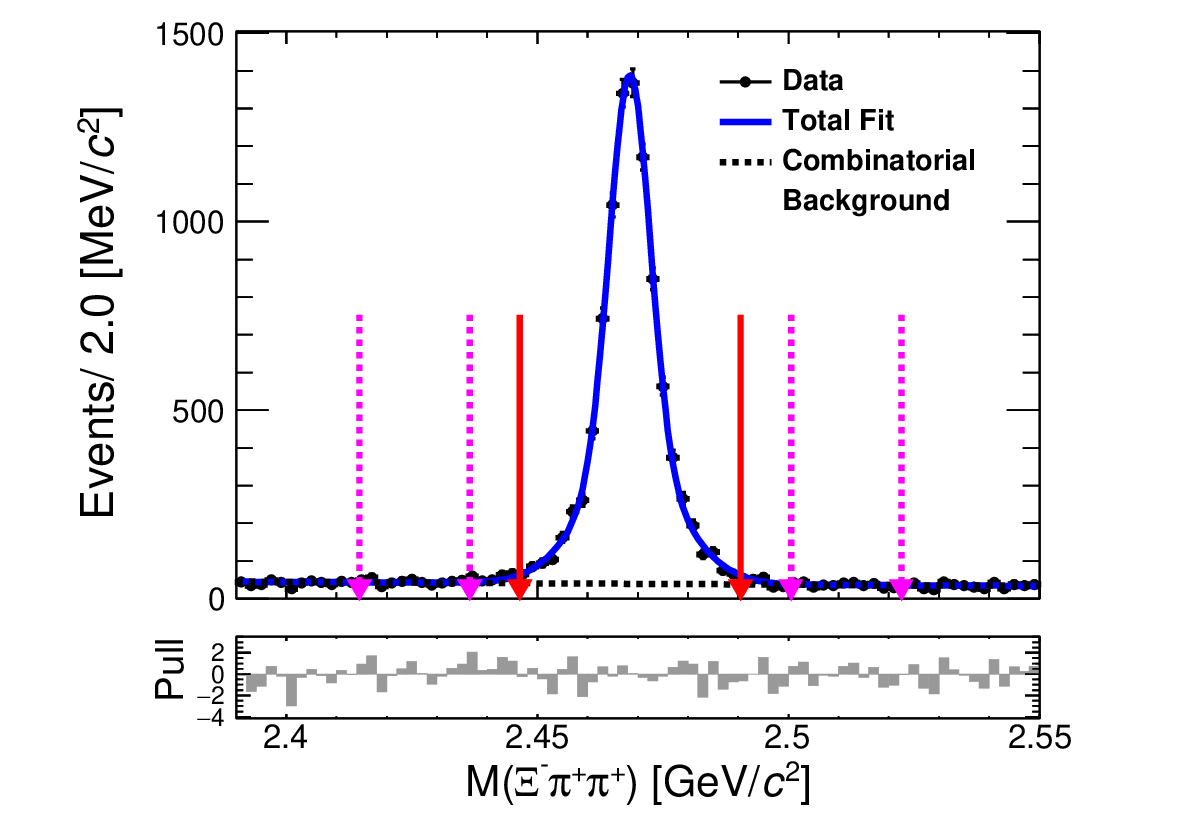}}	
		\end{minipage}
		\put(-380,65){\bf (a)} \put(-165,65){\bf (b)} 
		\put(-380,80){\scriptsize Belle} \put(-355,80){\scriptsize $\int \lum dt$ = 983.0~fb$^{-1}$}
		\put(-165,80){\scriptsize Belle~II} \put(-130,80){\scriptsize $\int \lum dt$ = 427.9~fb$^{-1}$}
	    \put(-380,50){\textcolor{gray}{preliminary}} \put(-165,50){\textcolor{gray}{preliminary}} 	
		\caption{The invariant mass spectra of $\Xi^- \pi^+ \pi^+$ from the reconstructed $\Xi_c^+ \to \Xi^- \pi^+ \pi^+$ 
		candidates in (a) Belle and (b) Belle~II data. The points with error bars represent the data, the solid blue
		curves show the best-fit results, and the dashed black curves represent the fitted combinatorial backgrounds.
		The solid red arrows indicate the defined $\Xi_c^+$ signal region: $|M(\Xi^- \pip \pip) - m_{\Xi_c^+}| < 22$~MeV/$c^2$($\sim$3$\sigma$),
		and the dashed magenta arrows denote the defined sideband regions: 
		$32~{\rm MeV}/c^2 < |M(\Xi^- \pip \pip) - m_{\Xi_c^+}| < 54~{\rm MeV}/c^2$.}\label{Fig2}
	\end{center}
\end{figure}

\noindent After applying all the selection criteria described above, the invariant mass spectra of $\Xi^- \pi^+ \pi^+$
from the reconstructed $\Xi_c^+ \to \Xi^- \pi^+ \pi^+$ candidates in Belle and Belle~II data are shown 
in figures~\ref{Fig2}(a) and \ref{Fig2}(b), respectively. To extract the yield of $\Xi_c^+ \to \Xi^- \pi^+ \pi^+$ signal events,
we perform an unbinned extended maximum-likelihood fit to the $M(\Xi^- \pi^+ \pi^+)$ distributions. In the fit,
a double-Gaussian function is used as the signal probability density function (PDF) for the $\Xi_c^+$ candidates,
while the combinatorial background PDF is parametrized by a second-order polynomial. All parameters of the signal and 
combinatorial background PDFs are free in the fit. The pull distributions, 
defined as $(N_{\rm data}-N_{\rm fit})/\sqrt{N_{\rm data}}$, 
are also shown in figure~\ref{Fig2}, where $N_{\rm data}$ represents the number of entries in each bin from data and 
$N_{\rm fit}$ is the number of events in each bin according to the fit.
The fitted $\Xi_c^+ \to \Xi^- \pi^+ \pi^+$ signal yields in Belle and Belle~II data are listed in table~\ref{tab2}.

\begin{figure}[htbp]
	\begin{minipage}{0.49\textwidth}
		\centerline{\includegraphics[width=8cm]{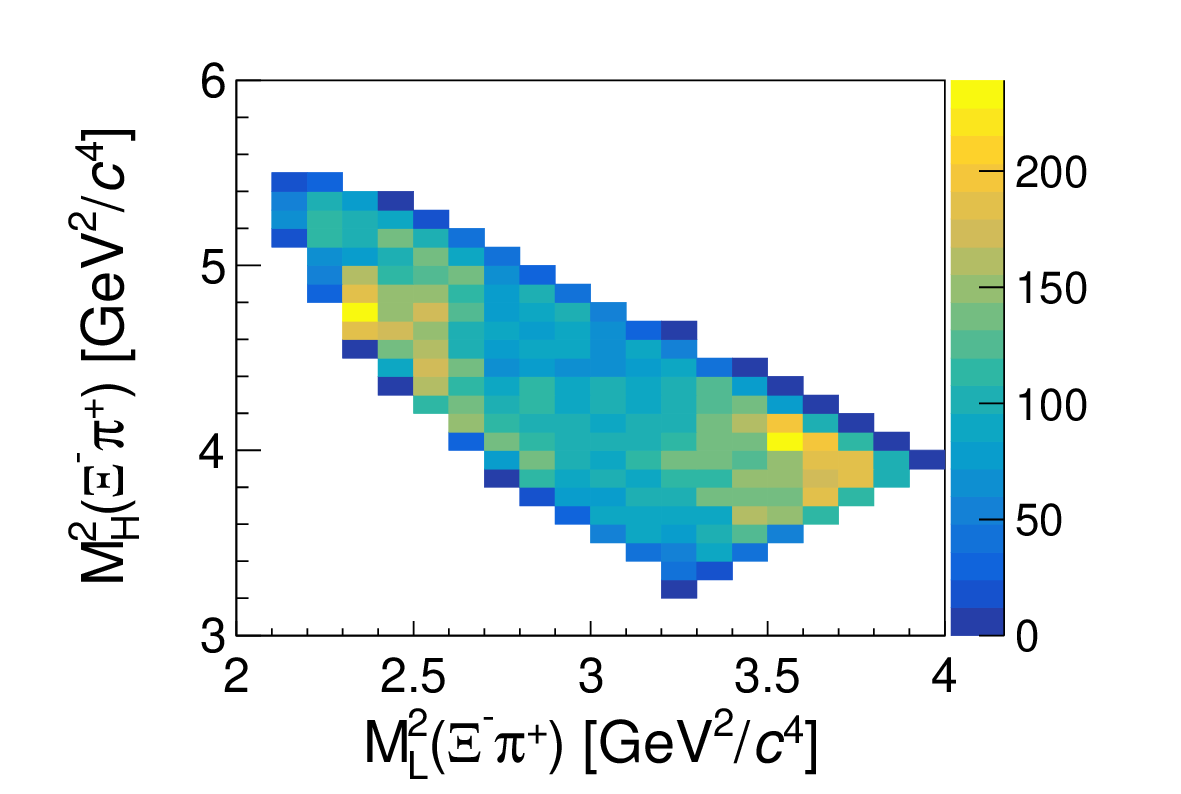}}
	\end{minipage}
	\begin{minipage}{0.49\textwidth}
		\centerline{\includegraphics[width=8cm]{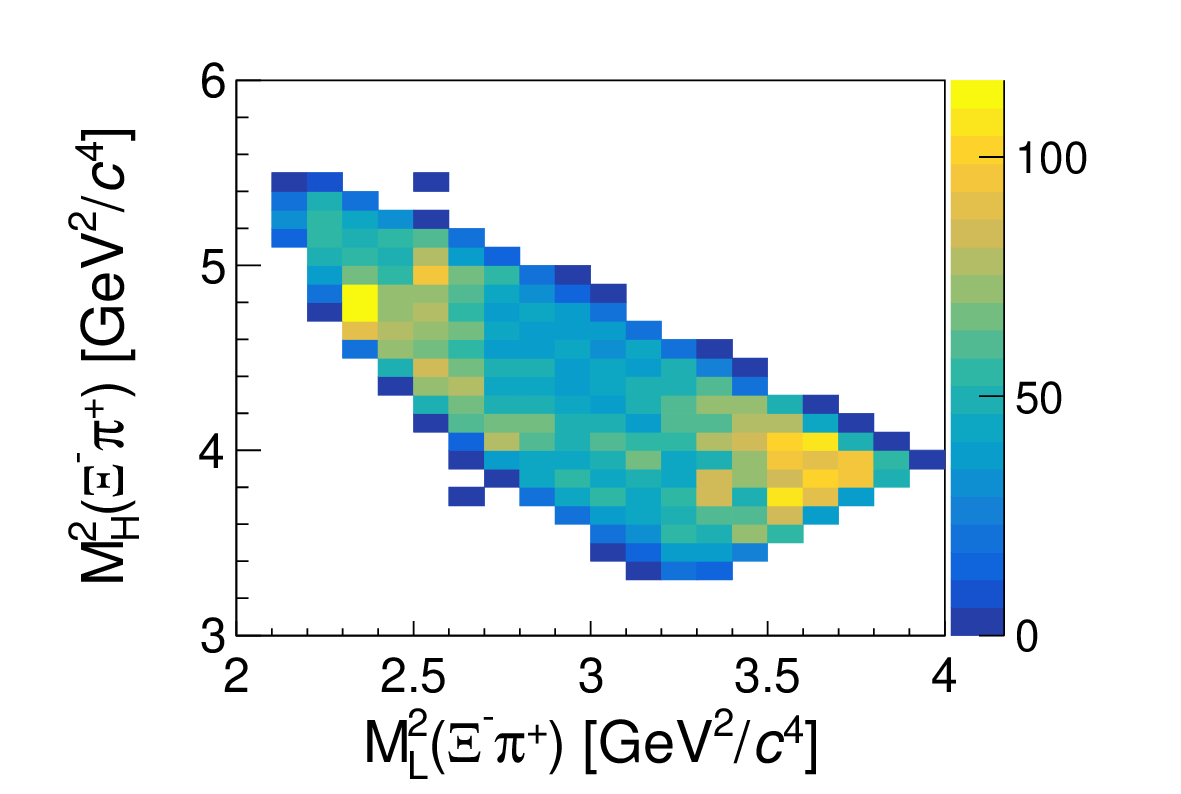}}	
	\end{minipage}
	\put(-274,50){\bf (a)} \put(-60,50){\bf (b)} 
	\put(-380,70){\scriptsize Belle} \put(-355,70){\scriptsize $\int \lum dt$ = 983.0~fb$^{-1}$}
	\put(-165,70){\scriptsize Belle~II} \put(-130,70){\scriptsize $\int \lum dt$ = 427.9~fb$^{-1}$}	
	\put(-380,52){\textcolor{gray}{preliminary}} \put(-165,52){\textcolor{gray}{preliminary}} 
	\caption{Dalitz plots of the reconstructed $\Xi_c^+ \to \Xi^{-}\pip\pip$ candidates from (a) Belle and (b) Belle II data 
	in the $\Xi_{c}^{+}$ signal region with the normalized $\Xi_c^+$ sideband events subtracted.}\label{Fig3}
\end{figure}
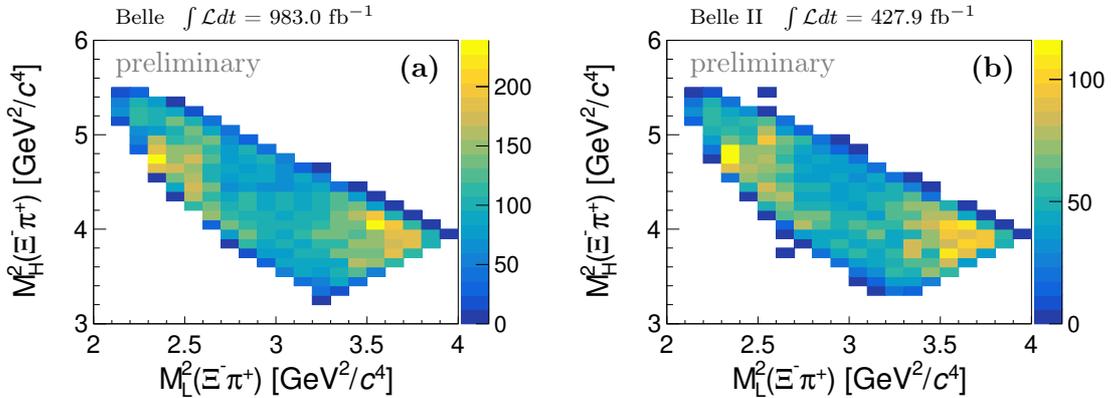

For the three-body decay $\Xi_c^+ \to \Xi^{-} \pip \pip$, the reconstruction efficiency can vary across
the phase space, as visualized in a Dalitz plot~\cite{Dalitz:1953cp}. 
Figures~\ref{Fig3}(a) and \ref{Fig3}(b) show the Dalitz plots of $M_{H}^2(\Xi^{-}\pi^{+})$ versus $M_{L}^2(\Xi^{-}\pi^{+})$
from Belle and Belle~II data in the $\Xi_{c}^{+}$ signal region, after subtracting the normalized $\Xi_c^+$ sideband events.
Here, the $\Xi^{-}\pi^{+}$ combination with a higher (lower) invariant mass is labeled as $M_{H}^2(\Xi^{-}\pi^{+})$ ($M_{L}^2(\Xi^{-}\pi^{+})$).
We divide the Dalitz plot into $20 \times 30$ bins and then apply a bin-by-bin correction for efficiency.
The reconstruction efficiency averaged over the Dalitz plot is determined by the formula 

\begin{equation}
\epsilon^{\rm corr} = \frac{\displaystyle \sum_{i} N_{s,i}}{\displaystyle \sum_{j} \left( \frac{N_{s,j}}{\epsilon_{j}} \right)},
\end{equation}
where $i$ and $j$ index all bins; $N_{s,i(j)}$ denotes the number of signal events in the $i(j)^{\rm th}$-bin in data;  
$\epsilon_{j}$ is the reconstruction efficiency from signal MC simulation for the $j^{\rm th}$-bin.
The term $N_{s,i}$ is calculated using $N_{i}^{\rm tot} - N^{\rm bkg}_{\rm SR}f_{i}^{\rm bkg}$, 
where $N_{i}^{\rm tot}$ is the number of total events in the $i^{\rm th}$-bin of the Dalitz plot
in data, $N^{\rm bkg}_{\rm SR}$ is the number of fitted background 
events in the $\Xi_{c}^{+}$ signal region in data, and $f_{i}^{\rm bkg}$ is the fraction
of background in the $i^{\rm th}$-bin, with $\Sigma_{i}f_{i}^{\rm bkg} = 1$. 
These fractions are obtained from the Dalitz plot of events in the
normalized $\Xi_{c}^{+}$ sideband regions in data. 
The average reconstruction efficiencies for the $\Xi_{c}^+ \to \Xi^- \pi^+ \pi^+$ decay 
are listed in table~\ref{tab2}.

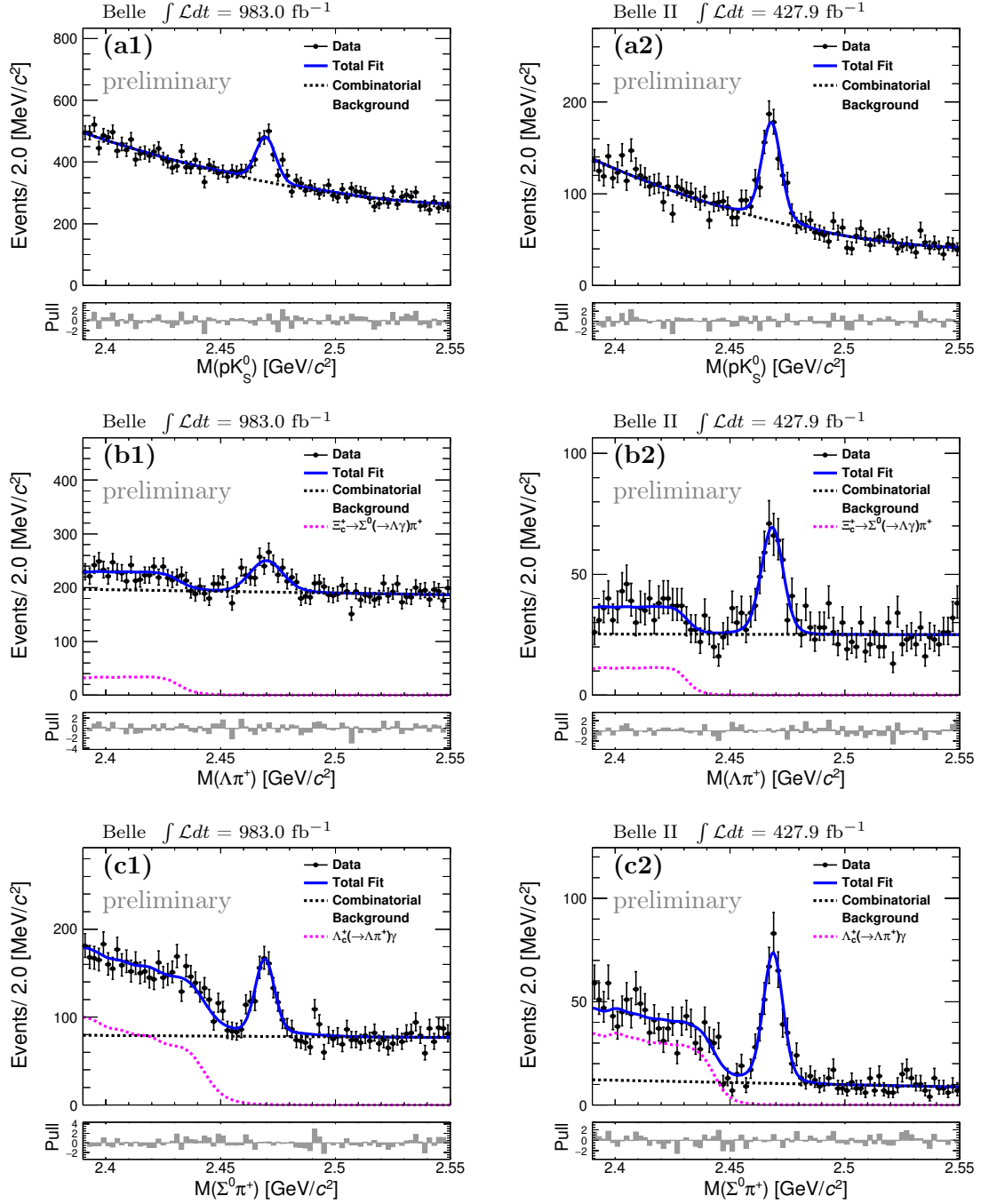
\begin{figure}[htbp]
	\begin{center}
		\begin{minipage}{0.49\textwidth}
			\centerline{\includegraphics[width=8cm]{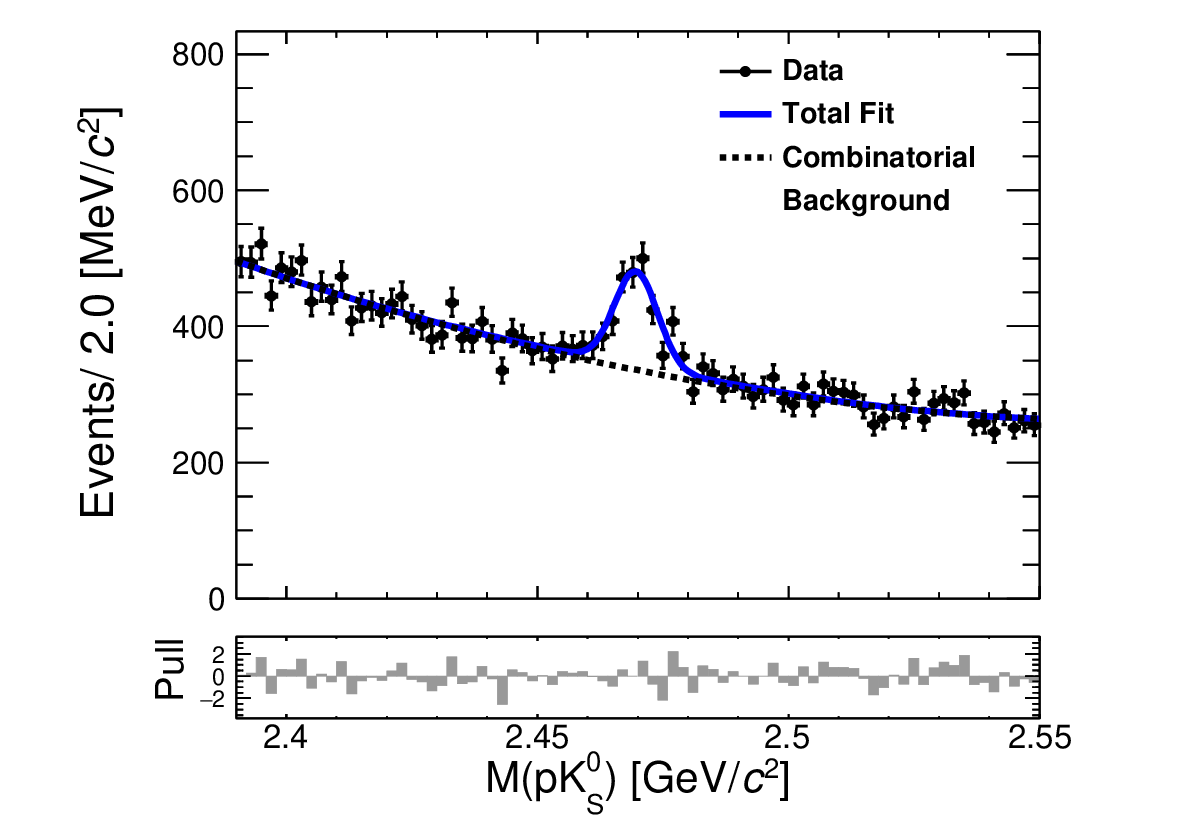}}
		\end{minipage}
		\begin{minipage}{0.49\textwidth}
			\centerline{\includegraphics[width=8cm]{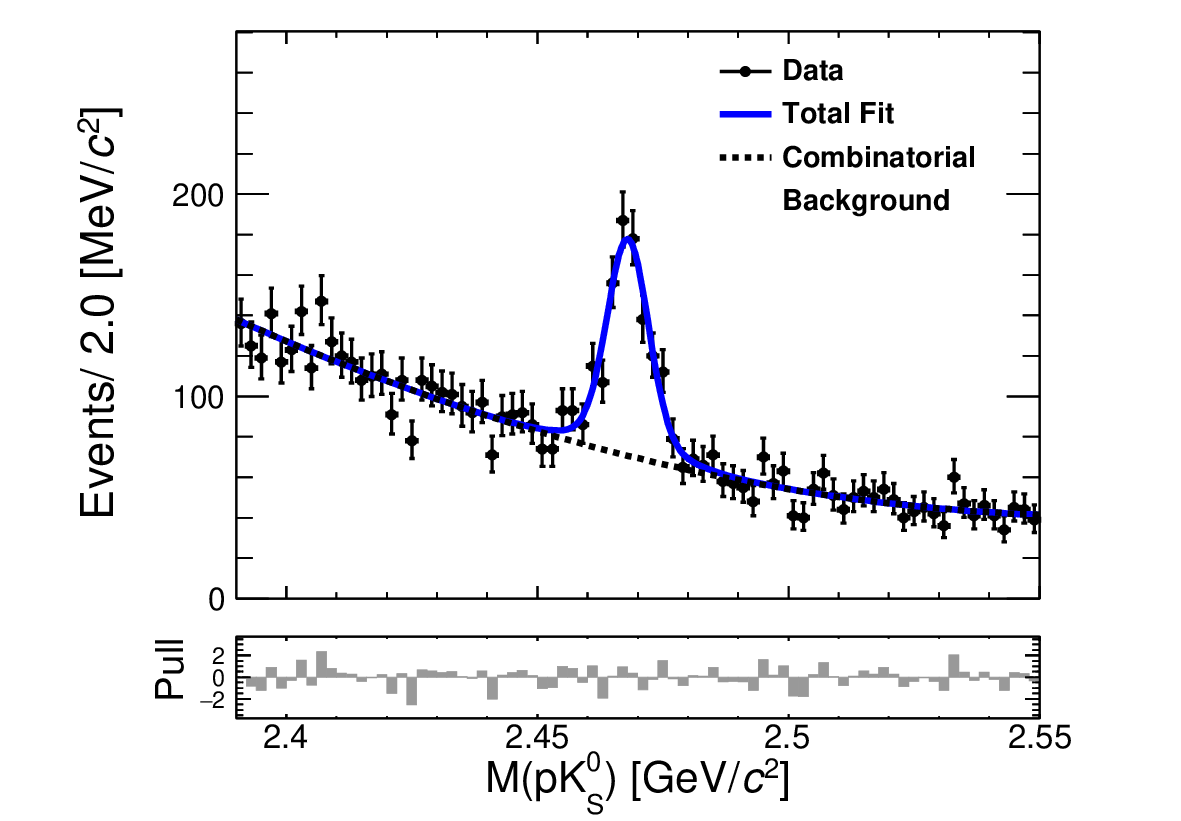}}	
		\end{minipage}
		\put(-380,65){\bf (a1)} \put(-165,65){\bf (a2)} 
		\put(-380,80){\scriptsize Belle} \put(-355,80){\scriptsize $\int \lum dt$ = 983.0~fb$^{-1}$}
		\put(-165,80){\scriptsize Belle~II} \put(-130,80){\scriptsize $\int \lum dt$ = 427.9~fb$^{-1}$}
		\put(-380,50){\textcolor{gray}{preliminary}} \put(-165,50){\textcolor{gray}{preliminary}} 	
		\vspace{0.35cm}

		\begin{minipage}{0.49\textwidth}
			\centerline{\includegraphics[width=8cm]{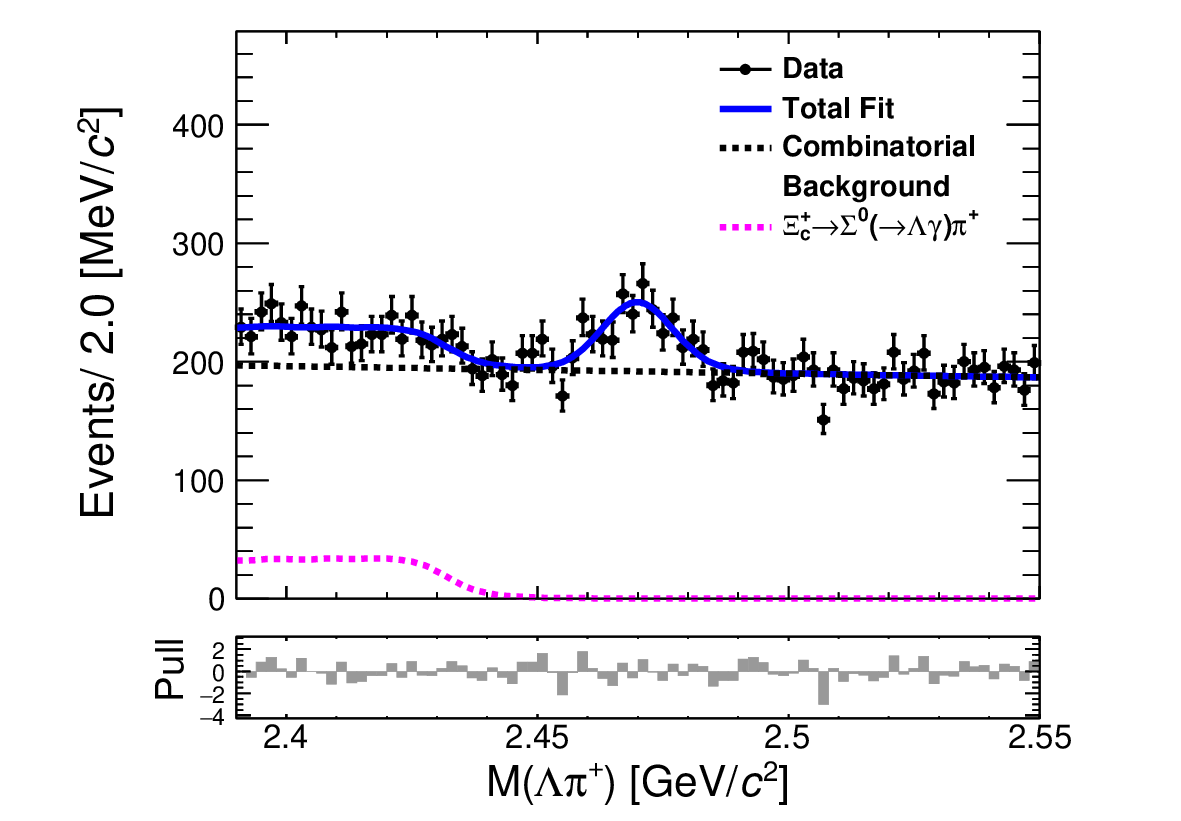}}
		\end{minipage}
		\begin{minipage}{0.49\textwidth}
			\centerline{\includegraphics[width=8cm]{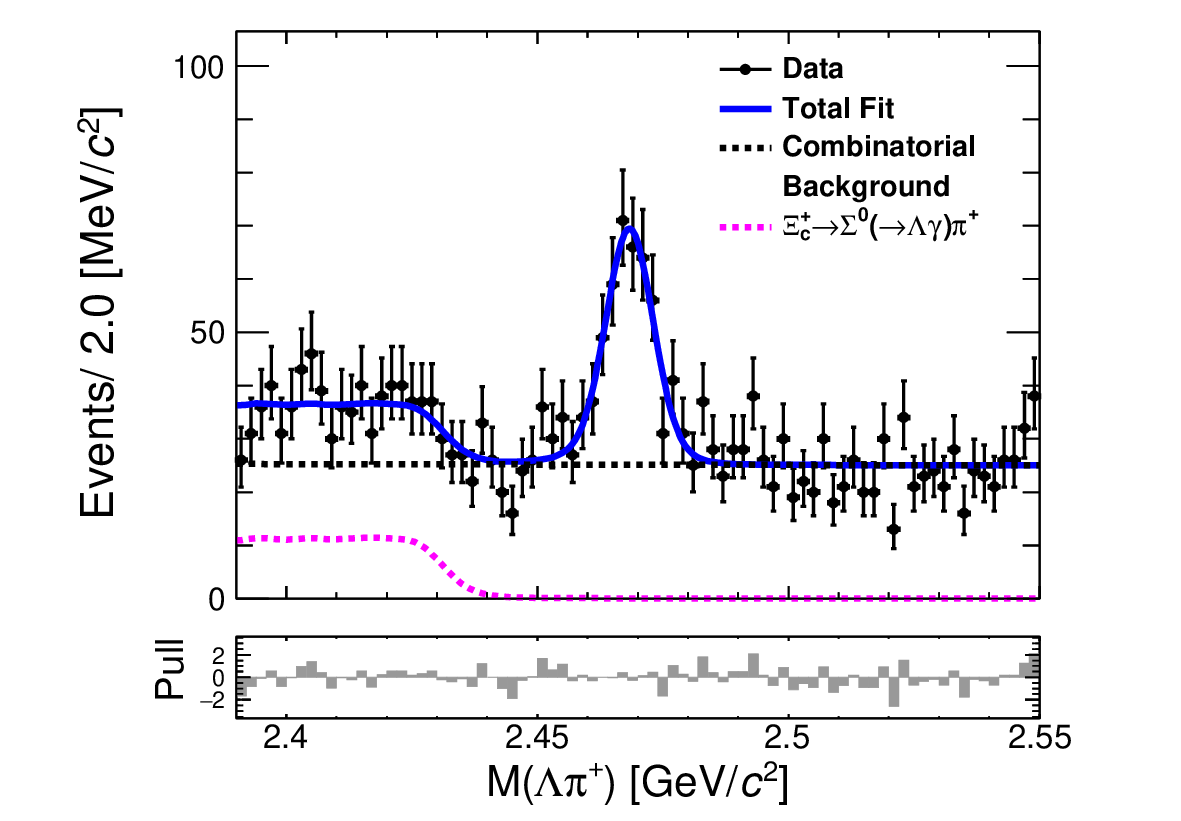}}	
		\end{minipage}
		\put(-380,65){\bf (b1)} \put(-165,65){\bf (b2)} 
		\put(-380,80){\scriptsize Belle} \put(-355,80){\scriptsize $\int \lum dt$ = 983.0~fb$^{-1}$}
		\put(-165,80){\scriptsize Belle~II} \put(-130,80){\scriptsize $\int \lum dt$ = 427.9~fb$^{-1}$}
		\put(-380,50){\textcolor{gray}{preliminary}} \put(-165,50){\textcolor{gray}{preliminary}} 
		
		\vspace{0.35cm}
		
		\begin{minipage}{0.49\textwidth}
			\centerline{\includegraphics[width=8cm]{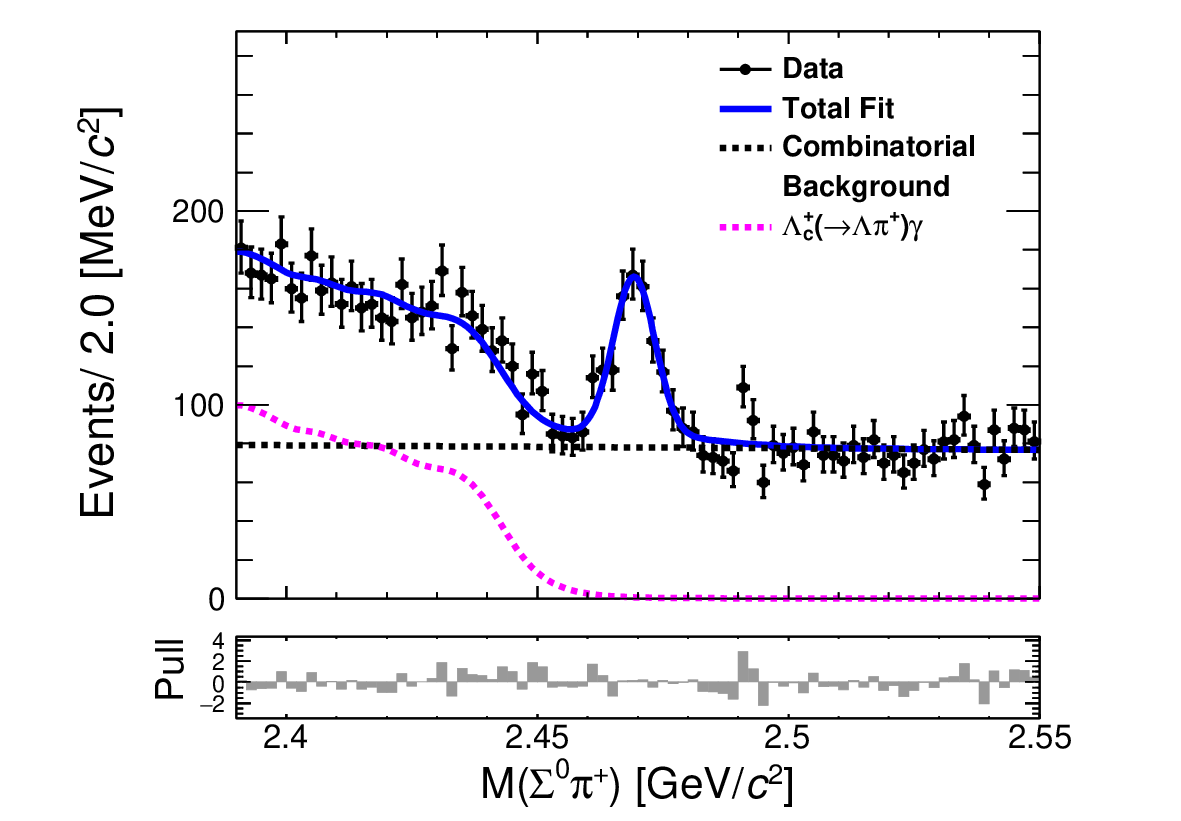}}
		\end{minipage}
		\begin{minipage}{0.49\textwidth}
			\centerline{\includegraphics[width=8cm]{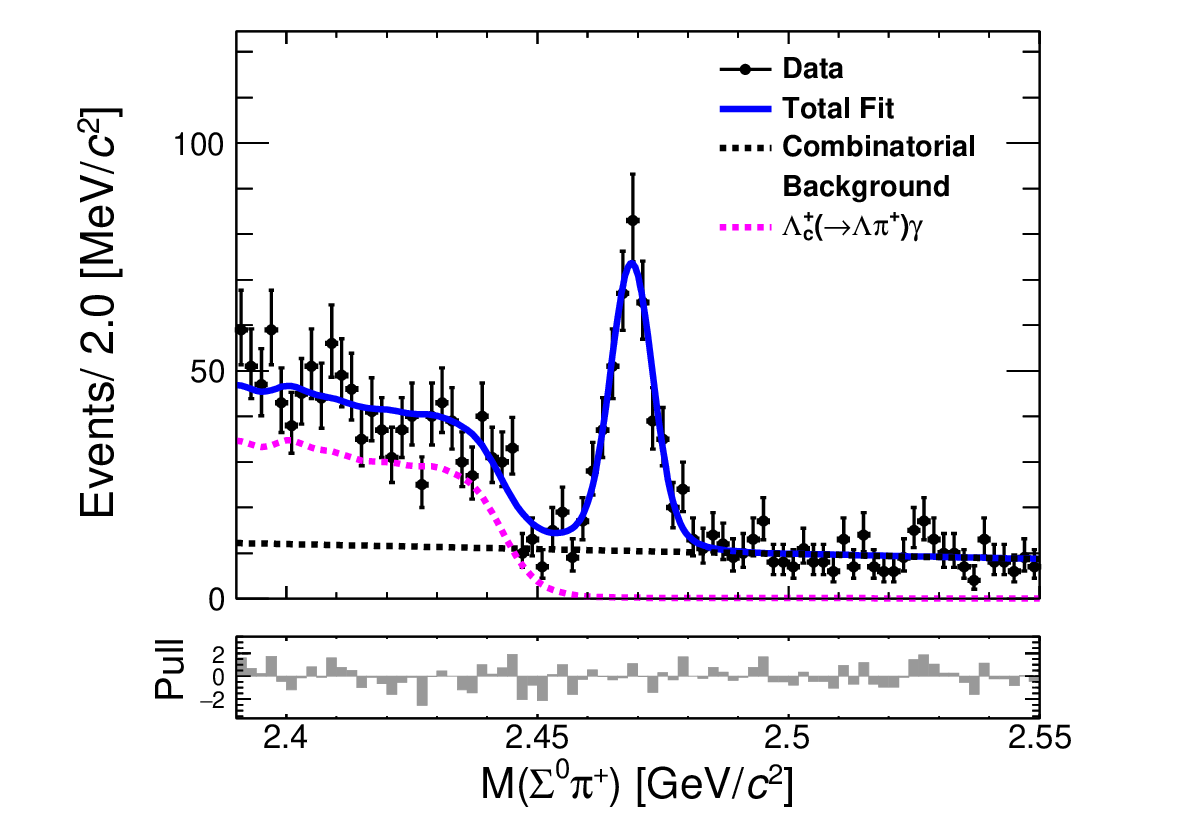}}	
		\end{minipage}
		\put(-380,65){\bf (c1)} \put(-165,65){\bf (c2)} 	
        \put(-380,80){\scriptsize Belle} \put(-355,80){\scriptsize $\int \lum dt$ = 983.0~fb$^{-1}$}
        \put(-165,80){\scriptsize Belle~II} \put(-130,80){\scriptsize $\int \lum dt$ = 427.9~fb$^{-1}$}
        \put(-380,50){\textcolor{gray}{preliminary}} \put(-165,50){\textcolor{gray}{preliminary}} 
		\caption{The invariant mass spectra of (a) $pK_S^0$, (b) $\Lambda \pi^+$, and (c) $\Sigma^0 \pi^+$ in
		(1) Belle and (2) Belle~II data. The points with error bars are the data, the solid blue curves
		show the best-fit results, and the dashed black curves represent the fitted combinatorial backgrounds.
		In the $M(\Lambda \pi^+)$ and $M(\Sigma^0 \pi^+)$ distributions, the dashed magenta curves are
		the fitted feed-down backgrounds from the $\Xi_c^+ \to \Sigma^0(\to \Lambda \gamma) \pi^+$ and 
		$\Lambda_c^+ \to \Lambda \pi^+$ decays, respectively.}\label{Fig4}
	\end{center}
\end{figure}

Figure~\ref{Fig4} shows the invariant mass spectra of $pK_S^0$, $\Lambda \pi^+$, and $\Sigma^0 \pi^+$ 
from Belle and Belle~II data. Significant signal yields are observed for all
three decays, in both Belle and Belle~II data. In addition, there are identified 
feed-down backgrounds with shoulder-like shapes on the lower mass sides of the $M(\Lambda \pi^+)$ and
$M(\Sigma^{0} \pi^+)$ distributions in both datasets. In the $M(\Lambda \pi^+)$ distribution, the 
feed-down background is attributed to the $\Xi_c^+ \to \Sigma^{0}(\to \Lambda \gamma) \pi^+$ decay
with a missing photon, while in the $M(\Sigma^0 \pi^+)$ distribution, it originates from a
$\Lambda_{c}^{+} \to \Lambda \pi^{+}$ decay combined with a random photon, as identified in the study 
of inclusive MC samples using the TopoAna package~\cite{topo}.

To extract the signal yields for $\Xi_c^+ \to pK_S^0$, $\Lambda \pi^+$, and $\Sigma^0 \pi^+$ in data, we
perform unbinned extended maximum-likelihood fits to the $M(pK_S^0)$, $M(\Lambda \pi^+)$, and $M(\Sigma^0 \pi^+)$ distributions. 
The signal shapes for 
$\Xi_c^+$ candidates are modeled by double-Gaussian functions with different mean values, and the fraction
and parameters of the tail Gaussian, which represents the broader part of the distribution, are fixed to
those obtained from the corresponding signal MC simulation. The combinatorial
backgrounds are parametrized by a second-order polynomial for the $M(pK_S^0)$ distribution, and first-order polynomials for the
$M(\Lambda \pi^+)$, and $M(\Sigma^0 \pi^+)$ distributions. The shapes of the feed-down backgrounds are represented 
by nonparametric (multi-dimensional) kernel-estimated probability density functions~\cite{Cranmer:2000du},
derived from the signal MC simulations. The fit results are displayed in figure~\ref{Fig4} 
along with the pull distributions, and the fitted signal yields are summarized in table~\ref{tab2}. 
The statistical significances of all decay channels are greater than 10$\sigma$ in both Belle and Belle II data, 
except for the $\Xi_c^+ \to \Lambda \pi^+$ decay in Belle data, which has a statistical significance of 7.6$\sigma$.
These significances are calculated using $-2\ln(\mathcal{L}_{0}/\mathcal{L}_\text{max})$~\cite{Wilks:1938dza},
accounting for the difference in the number of degrees of freedom ($\Delta {\rm ndf} = 3$),  where $\mathcal{L}_{0}$ 
and $\mathcal{L}_\text{max}$  are the maximized likelihoods without and with a signal component, respectively.

\begin{table}[htbp]
	\begin{center}
	\centering
	\caption{Summary of the fitted signal yields ($N^{\rm fit}$) and reconstruction efficiencies ($\epsilon$).
	All the uncertainties here are statistical only. The efficiencies for Belle II are higher than those for Belle.
    This is mostly due to the different requirements for the significance of the flight distance of $\Xi_c^+$. 
    This improvement is due to the superior vertex resolution of the Belle II VXD detector and the smaller beam spot
    of SuperKEKB, enabling Belle~II to achieve a larger efficiency while effectively excluding more backgrounds.} 
	\vspace{0.2cm}
	\label{tab2}
	\begin{tabular}{|l@{\hskip 10pt}r@{\hskip 10pt}c@{\hskip 12pt}cc|}
	\hline
	\multirow{2}{*}{Mode} & \multicolumn{2}{c}{Belle} & \multicolumn{2}{c}{Belle~II} \vline \\
	& \multicolumn{1}{c}{$N^{\text{fit}}$} & \multicolumn{1}{c}{$\epsilon$ (\%)}  & \multicolumn{1}{c}{$N^{\text{fit}}$} & \multicolumn{1}{c}{ $\epsilon$ (\%)} \vline \\
	\hline
	$\Xi_c^+ \to \Xi^- \pi^+ \pi^+$   & $17657 \pm 160$                      & $3.63 \pm 0.03$  & $8970 \pm 106$ & $\phantom{1}4.61  \pm 0.03$  \\
    $\Xi_{c}^{+}\to p K_S^0$          & $\phantom{2}917 \pm 103$             & $7.39 \pm 0.04$  & $\phantom{1}608 \pm \phantom{1}45$   & $11.30 \pm 0.04$ \\
    $\Xi_{c}^{+}\to \Lambda\pi^{+}$   & $\phantom{2}530 \pm \phantom{1}88$   & $6.35 \pm 0.04$  & $\phantom{1}275 \pm \phantom{1}30$   & $\phantom{1}9.62  \pm 0.03$  \\
    $\Xi_{c}^{+}\to\Sigma^{0}\pi^{+}$ & $\phantom{2}537 \pm \phantom{1}57$   & $2.78 \pm 0.03$  & $\phantom{1}359 \pm \phantom{1}27$   & $\phantom{1}4.33  \pm 0.03$  \\
	\hline
    \end{tabular}
    \end{center}
\end{table}

The ratios of branching fractions of $\Xi_{c}^{+} \to p K_S^0/\Lambda\pi^+/\Sigma^0\pip$ 
relative to that of $\Xi_{c}^{+} \to \Xi^- \pip \pip$ are calculated separately for Belle and Belle~II 
data using the formulas

\begin{equation}
	\frac{\BR(\Xi_{c}^+ \to p K_S^0)}{\BR(\Xi_c^+ \to \Xi^- \pip \pip)} = \frac{N^{\rm fit}(\Xi_{c}^+ \to p K_S^0)\eff(\Xi_c^+ \to \Xi^- \pip \pip)\BR(\Xi^- \to \Lambda \pim)\BR(\Lambda \to p \pim)}{N^{\rm fit}(\Xi_{c}^+ \to \Xi^- \pip \pi^+)\eff(\Xi_c^+ \to p K_S^0)\BR(K_S^0 \to \pip \pim)},
\end{equation}

\begin{equation}
	\frac{\BR(\Xi_{c}^+ \to \Lambda \pip)}{\BR(\Xi_c^+ \to \Xi^- \pip \pip)} = \frac{N^{\rm fit}(\Xi_{c}^+ \to \Lambda \pip)\eff(\Xi_c^+ \to \Xi^- \pip \pip)\BR(\Xi^- \to \Lambda \pim)}{N^{\rm fit}(\Xi_{c}^+ \to \Xi^- \pip \pi^+)\eff(\Xi_c^+ \to \Lambda \pip)},
\end{equation}
and

\begin{equation}
	\frac{\BR(\Xi_{c}^+ \to \Sigma^{0} \pip)}{\BR(\Xi_c^+ \to \Xi^- \pip \pip)} = \frac{N^{\rm fit}(\Xi_{c}^+ \to \Sigma^{0} \pip)\eff(\Xi_c^+ \to \Xi^- \pip \pip)\BR(\Xi^- \to \Lambda \pim)}{N^{\rm fit}(\Xi_{c}^+ \to \Xi^- \pip \pi^+)\eff(\Xi_c^+ \to \Sigma^{0} \pip)\BR(\Sigma^{0} \to \Lambda \gamma)},
\end{equation}
where $N^{\rm fit}(\Xi_{c}^+ \to p K_S^0)$, $N^{\rm fit}(\Xi_{c}^+ \to \Xi^- \pip \pip)$, $N^{\rm fit}(\Xi_{c}^+ \to \Lambda \pi^+)$, 
and $N^{\rm fit}(\Xi_{c}^+ \to \Sigma^0 \pi^+)$ are the numbers of fitted $\Xi_{c}^+ \to p K_S^0$, $\Xi_c^+ \to \Xi^- \pip \pip$, 
$\Xi_{c}^+ \to \Lambda \pi^+$, and $\Xi_{c}^+ \to \Sigma^0 \pi^+$ signal events in data summarized in table~\ref{tab2}; 
$\eff(\Xi_c^+ \to p K_S^0)$, $\eff(\Xi_c^+ \to \Xi^- \pip \pip)$, $\eff(\Xi_{c}^+ \to \Lambda \pi^+)$, 
and $\eff(\Xi_{c}^+ \to \Sigma^0 \pi^+)$ are the corresponding reconstruction efficiencies listed in table~\ref{tab2};
and $\BR(\Xi^- \to \Lambda \pim)$ = (99.887 $\pm$ 0.035)\%, $\BR(\Lambda \to p \pim)$ = (64.2 $\pm$ 0.5)\%, 
$\BR(K_S^0 \to \pip \pim)$ = (69.20 $\pm$ 0.05)\%, and $\BR(\Sigma^{0} \to \Lambda \gamma) = 100\%$ 
are taken from the Particle Data Group~\cite{ParticleDataGroup:2024cfk}. 
The reconstruction efficiencies of signal channels are obtained from simulation using
the ratio $N_{\rm sel.}/N_{\rm gen.}$, where $N_{\rm sel.}$ and $N_{\rm gen.}$ are the numbers of true 
signal events surviving the selection criteria and generated events, respectively.
The calculated branching fraction ratios are summarized in table~\ref{tab3}.

We combine the ratios of branching fractions and uncertainties measured at Belle and Belle~II using the formulas in ref.~\cite{DAgostini:1993arp}

\begin{equation}\label{combined}
	r=\frac{r_{1}\sigma_{2}^{2}+r_{2}\sigma_{1}^{2}}{\sigma_{1}^{2}+\sigma_{2}^{2}+(r_{1}-r_{2})^{2}\epsilon_{r}^{2}},
\end{equation}
and 

\begin{equation}\label{combined2}
	\sigma = \sqrt{\frac{\sigma_{1}^{2}\sigma_{2}^{2}+(r_{1}^{2}\sigma_{2}^{2}+r_{2}^{2}\sigma_{1}^{2})\epsilon_{r}^{2}}{\sigma_{1}^{2}+\sigma_{2}^{2}+(r_{1}-r_{2})^{2}\epsilon_{r}^{2}}}
\end{equation}
where $r_{i}$, $\sigma_{i}$, and $\epsilon_{r}$ are the branching fraction ratio, uncorrelated uncertainty, and relative
correlated systematic uncertainty from each data sample, respectively. The correlated systematic uncertainty
includes branching fractions of intermediate states and the background shape in the fit, while the efficiency-related 
uncertainty is treated as uncorrelated. All the uncorrelated and correlated uncertainties are listed in table~\ref{tab4}.
The combined branching fraction ratios are summarized 
in table~\ref{tab3}, where the first and second uncertainties are statistical and systematic, respectively.
The systematic uncertainties are discussed in detail below.

\begin{table}[htbp]
	\centering
	\caption{The ratios of branching fractions of $\Xi_{c}^{+} \to p K_S^0/\Lambda\pi^+/\Sigma^0\pi^+$ relative to that of $\Xi_{c}^{+} \to \Xi^- \pi^+ \pi^+$, 
		where the first and second uncertainties are statistical and systematic, respectively.}
	\vspace{0.2cm}
	\label{tab3}
	\begin{tabular}{|lccc|}
		\hline
		& Belle & Belle~II & Combined \\
		\hline
		$\frac{\BR(\Xi_{c}^{+}\to p K_S^0)}{\BR(\Xi_c^+ \to \Xi^- \pi^+ \pi^+)}$  & $(2.36 \pm 0.27 \pm 0.08)\%$  & $(2.56 \pm 0.19 \pm 0.11)\%$  & $(2.47 \pm 0.16 \pm 0.07)\%$   \\
		$\frac{\BR(\Xi_{c}^{+}\to \Lambda\pi^{+})}{\BR(\Xi_c^+ \to \Xi^- \pi^+ \pi^+)}$   & $(1.72 \pm 0.29 \pm 0.11)\%$  & $(1.47 \pm 0.16 \pm 0.09)\%$  & $(1.56 \pm 0.14 \pm 0.09)\%$  \\
		$\frac{\BR(\Xi_{c}^{+}\to\Sigma^{0}\pi^{+})}{\BR(\Xi_c^+ \to \Xi^- \pi^+ \pi^+)}$ & $(3.97 \pm 0.42 \pm 0.23)\%$  & $(4.26 \pm 0.33 \pm 0.24)\%$  & $(4.13 \pm 0.26 \pm 0.22)\%$ \\
		\hline
	\end{tabular}
\end{table}

\section{Systematic uncertainties} 
\noindent Sources of systematic uncertainties in the measurements of 
the branching fraction ratios include those associated with efficiency, 
the branching fractions of intermediate states, and the fit procedure.
Note that some uncertainties from efficiency-related sources and the branching
fractions of intermediate states cancel when taking the ratio to the normalization mode.
Table~\ref{tab4} summarizes these systematic uncertainties, with the total 
uncertainty calculated as the quadratic sum of the uncertainties from each source.

\begin{table}[htbp]
	\centering
	\caption{\label{tab4} Relative systematic uncertainties (\%) in the measurements of branching fraction ratios.
	The uncertainties due to intermediate branching fractions and fit uncertainty
	are common to Belle and Belle II; the other uncertainties are independent.}
	\vspace{0.2cm}
	\begin{tabular}{|lcccccc|}
		\hline
		\multirow{2}{*}{Sources} &  \multicolumn{2}{c}{$\frac{\BR(\Xi_{c}^{+} \to p K_{S}^{0})}{\BR(\Xi_{c}^{+} \to \Xi^{-}  \pi^{+} \pi^{+})}$} & 
		\multicolumn{2}{c}{$\frac{\BR(\Xi_{c}^{+} \to \Lambda \pi^+)}{\BR(\Xi_{c}^{+} \to \Xi^{-}  \pi^{+} \pi^{+})}$} &
		\multicolumn{2}{c|}{$\frac{\BR(\Xi_{c}^{+} \to \Sigma^0 \pi^+)}{\BR(\Xi_{c}^{+} \to \Xi^{-}  \pi^{+} \pi^{+})}$} \\ 
		
		&  Belle &  Belle~II   &  Belle &  Belle~II & Belle &  Belle~II           \\
		\hline
		Tracking                     &  0.7 &  0.7  &  0.7  &  0.7  &  0.7  &   0.7  \\
		Particle identification      &  0.1 &  0.2  &  0.1  &  0.1  &  0.1  &   0.1  \\
		$K_{S}^0$ reconstruction     &  0.8 &  2.6  &  ...  &  ...  &  ...  &   ...  \\
		$\Lambda$ reconstruction     &  0.5 &  0.3  &  0.3  &  0.2  &  0.3  &   0.2  \\
		Photon reconstruction        &  ... &  ...  &  ...  &  ...  &  2.0  &   1.1  \\
		Mass resolution              &  0.2 &  0.2  &  0.4  &  0.5  &  0.7  &   0.8  \\
		Dalitz efficiency correction &  1.3 &  1.5  &  1.3  &  1.5  &  1.3  &   1.5  \\
		Branching fraction           &  0.8 &  0.8  &  0.0  &  0.0  &  0.0  &   0.0  \\
       	Fit Uncertainty              &  2.5 &  2.5  &  5.9  &  5.9  &  5.1  &   5.1  \\
      	Sum in quadrature            &  3.2 &  4.1  &  6.1  &  6.2  &  5.7  &   5.5  \\  
		\hline
	\end{tabular}
\end{table} 

The systematic uncertainty of the efficiency determination includes effects due to the detection efficiency of
the daughter particles, the mass window used for the intermediate state, and the averaging
of the efficiency across the Dalitz plot of the normalization mode.
Based on the table of the detection efficiency ratios between data and MC ($r_{\epsilon}=\epsilon_{\rm data}/\epsilon_{\rm MC}$)
from the control sample, we build 1000 $r_{\epsilon}$ tables for both the signal and normalization modes by
randomly fluctuating $r_{\epsilon}$ in each bin according to its uncertainty and calculate $\bar{r}_{\epsilon}$ for each. 
We take the mean values from the distributions of $\bar{r}_{\epsilon}^{\rm sig.}$ and $\bar{r}_{\epsilon}^{\rm nor.}$ as the
efficiency correction factors of the signal and normalization modes, respectively, and the root-mean-square value from the distribution of
$\bar{r}_{\epsilon}^{\rm sig.}$/$\bar{r}_{\epsilon}^{\rm nor.}$ as the systematic uncertainty in the measurement of the
branching fraction ratio.
The efficiency correction factors and uncertainties include those from track-finding efficiency, obtained from
the control samples of $D^{*+} \to D^{0}(\to K_S^0 \pi^+ \pi^-)\pi^+$
at Belle and $\bar{B}^{0} \to D^{*+}(\to D^0\pi^+)\pi^-$ and $e^+e^- \to \tau^+ \tau^-$
at Belle~II; charged pion identification, obtained from the 
control samples of $D^{*+} \to D^{0}(\to K^- \pi^+)\pi^+$ at Belle and $K_S^0 \to \pi^+ \pi^-$
at Belle~II; proton identification, obtained from the $\Lambda \to p \pi^-$
control sample at Belle and Belle~II; $K_S^0$ reconstruction, obtained from the  control samples of 
$D^{*+} \to D^{0}(\to K_S^0 \pi^0)\pi^+$ at Belle and $D^{*+} \to D^{0}(\to K_S^0 \pi^+ \pi^-)\pi^+$ at Belle~II;
$\Lambda$ reconstruction, obtained for the control samples of $\Lambda \to p \pi^-$ at Belle 
and $\Lambda_c^+ \to \Lambda(\to p \pi^-) \pi^+$ at Belle~II; and photon reconstruction, obtained from control samples 
of radiative Bhabhas at Belle and radiative muon-pairs at Belle II. 
The PID uncertainties listed in table~\ref{tab4} include the proton and two-pion identification uncertainties 
in the measurement of $\BR(\Xi_c^+ \to pK_S^0)/\BR(\Xi_c^+ \to \Xi^- \pi^+ \pi^+)$ and three-pion identification uncertainties
in the measurements of $\BR(\Xi_c^+ \to \Lambda \pi^+)/\BR(\Xi_c^+ \to \Xi^- \pi^+ \pi^+)$ and 
$\BR(\Xi_c^+ \to \Sigma^0 \pi^+)/\BR(\Xi_c^+ \to \Xi^- \pi^+ \pi^+)$.
For the $\Lambda$ reconstruction, the momentum distributions of 
$\Lambda$ of the signal modes $\Xi_c^+ \to \Lambda \pi^+$ and $\Xi_c^+ \to \Sigma^0 \pi^+$ 
and the normalization mode overlap in most regions, but there are still some differences.
Therefore, we treat the efficiency correction factors for the signal and normalization modes separately, 
along with a
systematic uncertainty that includes the proton identification uncertainty, using the same method mentioned above.
The uncertainty due to the mass window requirement for the intermediate state is calculated based 
on the difference between the selected signal fractions
in the simulation and data. For the reference mode $\Xi_c^+ \to \Xi^- \pi^+ \pi^+$, the signal efficiency is corrected across the Dalitz plot. The selected $\Xi_{c}^{+}$ sideband regions may influence the efficiency. To account for this, we shift the $\Xi_c^+$ sideband regions by $\pm5$~MeV/$c^2$, and the average deviation 
in efficiency compared to the nominal value is taken as the systematic uncertainty. 
We assume that the decays $\Xi_c^+ \to pK_S^0$, $\Lambda \pi^+$, and $\Sigma^0 \pi^+$ are 
isotropic in the rest frame of the $\Xi_c^+$, and a phase space model is employed to generate signal MC events.
Since the efficiency-corrected $\cos\theta_{\text{hel}}$ distributions are consistent with those in the 
\mbox{MC signal} distributions at the generator level, where
$\theta_{\text{hel}}$ represents the helicity angle between the momentum of the daughter baryon $(p/\Lambda/\Sigma^0)$
and the opposite of the boost direction of the c.m.\ system, the systematic uncertainty associated with
the model of signal MC generation can be neglected. We weight the signal MC samples according to
the efficiency-corrected $x_p$ distribution of the normalization mode from data to ensure good agreement between data and MC.
The efficiency-corrected $x_p$ distribution is obtained by fitting the $M(\Xi^- \pi^+ \pi^+)$ distribution in each $x_p$ 
bin of data, while accounting for the efficiency in each bin.

For the measurement of $\BR(\Xi_{c}^+ \to p K_{S}^{0})/\BR(\Xi_{c}^+ \to \Xi^-\pi^+ \pi^+)$,
the uncertainties from $\BR(K_{S}^0 \to \pi^+ \pi^-)$, $\BR(\Xi^- \to \Lambda \pi^-)$, and 
$\BR(\Lambda \to p \pi^-)$ are 0.072\%, 0.035\%, and 0.78\%~\cite{ParticleDataGroup:2024cfk}, 
respectively. These uncertainties are combined in quadrature to obtain the total uncertainty 
from the branching fractions of intermediate states. For the measurements of $\BR(\Xi_{c}^+ \to \Lambda \pi^+)/\BR(\Xi_{c}^+ \to \Xi^-\pi^+ \pi^+)$ and 
$\BR(\Xi_{c}^+ \to \Sigma^{0} \pi^+)/\BR(\Xi_{c}^+ \to \Xi^-\pi^+ \pi^+)$, 
the uncertainty from $\BR(\Xi^- \to \Lambda \pi^-)$ is only 0.035\%~\cite{ParticleDataGroup:2024cfk}.

The systematic uncertainty associated with the fit procedure is evaluated by changing the background PDF 
to a higher-order polynomial or a lower-order polynomial, and the average deviation from the nominal
fit result is taken as the systematic uncertainty. Here, the uncertainty from the background
PDF is treated as correlated, and extracted from a simultaneous fit to Belle and Belle II data.
We estimate the fit uncertainties for both the signal and normalization modes separately, 
and the uncertainty for the normalization mode is determined to be 0.9\%. Finally, the fit uncertainties 
of the signal and normalization modes are added in quadrature to obtain the total fit uncertainty.
 
\section{Result and discussion}
\noindent 
In summary, we report the first observations of the singly Cabibbo-suppressed decays
$\Xi_c^{+} \to pK_{S}^{0}$, $\Xi_c^+ \to \Lambda \pi^+$, and $\Xi_c^+ \to \Sigma^{0} \pi^+$, 
each with a signal significance greater than 10$\sigma$, using the combined data samples of 983.0~$\rm fb^{-1}$
and 427.9~$\rm fb^{-1}$ collected by the Belle and Belle~II detectors. 
The ratios of branching fractions of the $\Xi_c^+ \to pK_S^0$, $\Lambda \pi^+$, and $\Sigma^0 \pi^+$ decays
relative to that of $\Xi_{c}^+ \to \Xi^- \pi^+ \pi^+$ are measured to be

\begin{equation}
	\frac{\BR(\Xi_c^+ \to pK_S^0)}{\BR(\Xi_c^{+} \to \Xi^{-} \pip \pip)} = (2.47 \pm 0.16 \pm 0.07)\% \notag,
\end{equation}

\begin{equation}
	\frac{\BR(\Xi_c^+ \to \Lambda \pi^+)}{\BR(\Xi_c^{+} \to \Xi^{-} \pip \pip)} = (1.56 \pm 0.14 \pm 0.09)\% \notag,
\end{equation}
and

\begin{equation}
	\frac{\BR(\Xi_c^+ \to \Sigma^0 \pi^+)}{\BR(\Xi_c^{+} \to \Xi^{-} \pip \pip)} = (4.13 \pm 0.26 \pm 0.22)\% \notag.
\end{equation}
Taking $\BR(\Xi_c^{+} \to \Xi^{-} \pip\pip) = (2.9 \pm 1.3)\%$~\cite{Belle:2019bgi},
the absolute branching fractions are determined to be

\begin{equation}
	\BR(\Xi_c^{+} \to p K_{S}^{0}) = (7.16 \pm 0.46 \pm 0.20 \pm 3.21) \times 10^{-4} \notag,
\end{equation}
\begin{equation}
	\BR(\Xi_c^{+} \to \Lambda \pip) = (4.52 \pm 0.41 \pm 0.26 \pm 2.03) \times 10^{-4} \notag,
\end{equation}
and

\begin{equation}
	\BR(\Xi_c^{+} \to \Sigma^0 \pip) = (1.20 \pm 0.08 \pm 0.07 \pm 0.54) \times 10^{-3} \notag.
\end{equation}
where the uncertainties are statistical, systematic, and from $\BR(\Xi_c^{+} \to \Xi^{-} \pip\pip)$, respectively.

\begin{figure}[htbp]
	\begin{minipage}{0.99\textwidth}
	\centerline{\hspace{4cm}\includegraphics[width=16cm]{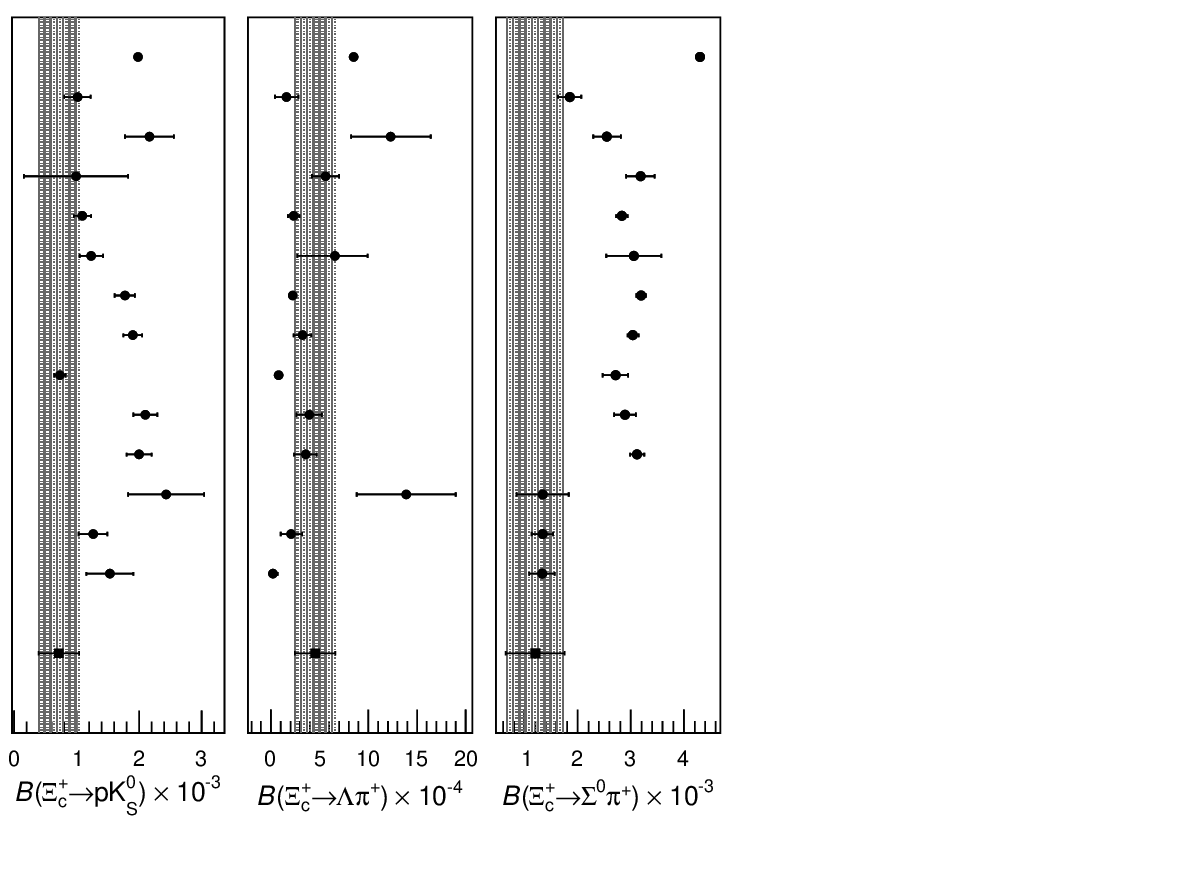}}
	\end{minipage}
   \put(-90,148){Zou {\it et.al}~\cite{Zou:2019kzq} }
   \put(-90,133){Geng {\it et.al}~\cite{Geng:2018plk} }
   \put(-90,117){Geng {\it et.al}~\cite{Geng:2019xbo} }
   \put(-90,102){Huang {\it et.al}~\cite{Huang:2021aqu} }
   \put(-90,87){Zhong {\it et.al} (I)~\cite{Zhong:2022exp} }
   \put(-90,71.5){Zhong {\it et.al} (II)~\cite{Zhong:2022exp} }
   \put(-90,56){Xing {\it et.al}~\cite{Xing:2023dni} }
   \put(-90,40){Geng {\it et.al}~\cite{Geng:2023pkr} }
   \put(-90,24.5){Liu~\cite{Liu:2023dvg} }
   \put(-90,9){Zhong {\it et.al} (I)~\cite{Zhong:2024qqs} }
   \put(-90,-6){Zhong {\it et.al} (II)~\cite{Zhong:2024qqs} }
   \put(-90,-21){Zhao {\it et.al}~\cite{Zhao:2018mov} }
   \put(-90,-36){Hsiao {\it et.al} (I)~\cite{Hsiao:2021nsc} }
   \put(-90,-51){Hsiao {\it et.al} (II)~\cite{Hsiao:2021nsc} }
   \put(-90,-77){Belle and Belle~II}
   \put(-90,-87){combined measurement}
   \put(-315,155){\bf (a)} \put(-220,155){\bf (b)} \put(-122,155){\bf (c)} 
   \caption{Comparisons of the measured (a) $\BR(\Xi_c^+ \to pK_S^0)$, (b) $\BR(\Xi_c^+ \to \Lambda \pi^+)$, and (c) $\BR(\Xi_c^+ \to \Sigma^0 \pi^+)$
   	with theoretical predictions~\cite{Zou:2019kzq, Zhao:2018mov,Geng:2018plk,Geng:2019xbo,Huang:2021aqu,Hsiao:2021nsc,Zhong:2022exp,Xing:2023dni,Geng:2023pkr,Liu:2023dvg,Zhong:2024qqs}.
    The dots and error bars represent the central values and uncertainties of the theoretical predictions, respectively. 
    The dots without error bars indicate that no theoretical uncertainty is available.
    The squares and error bars denote the measured central values and uncertainties in this work.
   	For refs.~\cite{Zhong:2022exp, Hsiao:2021nsc}, (I) indicates the 
   	predicted value based on the $\rm SU(3)_F$ flavor symmetry, while (II) takes into account the breaking of $\rm SU(3)_F$ flavor symmetry. 
   	For ref.~\cite{Zhong:2024qqs}, (I) and (II) represent the predicted values derived from the topological diagrammatic approach and the 
   	irreducible $\rm SU(3)_F$ approach, respectively.
}\label{Fig5}
\end{figure}
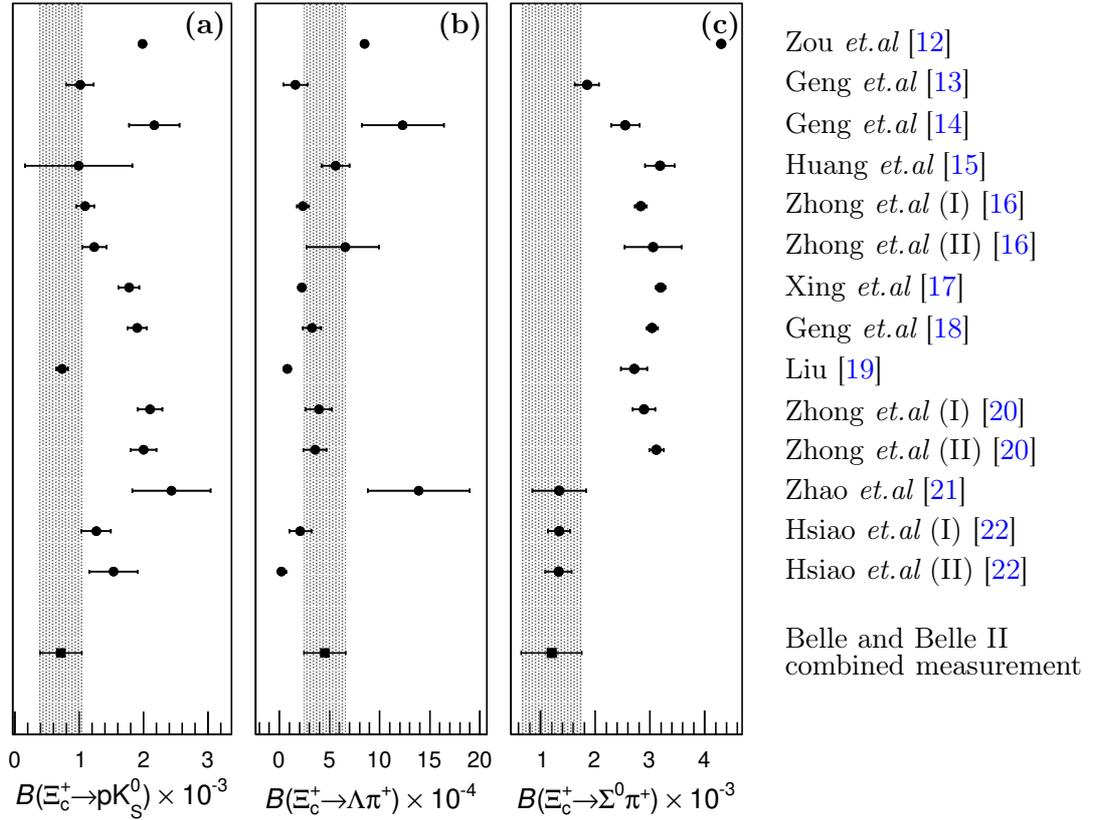

Figure~\ref{Fig5} presents comparisons of the measured absolute branching 
fractions of $\Xi_c^+ \to pK_S^0$, $\Lambda \pi^+$, and $\Sigma^0 \pi^+$ decays 
in this work with the theoretical predictions~\cite{Zou:2019kzq, Zhao:2018mov,Geng:2018plk,Geng:2019xbo,Huang:2021aqu,Hsiao:2021nsc,Zhong:2022exp,Xing:2023dni,Geng:2023pkr,Liu:2023dvg,Zhong:2024qqs}.
The $\chi^2$ values for the predicted branching fractions in refs.~\cite{Geng:2018plk,Hsiao:2021nsc}
compared to the experimental measured results for each of these three decay modes are all less than 4.
The measured absolute branching fractions of $\Xi_c^+ \to pK_S^0$ and $\Xi_c^+ \to \Sigma^0 \pi^+$ 
are lower than the central values predicted by most theoretical papers. 
However, the measured absolute branching fraction of $\Xi_c^+ \to \Lambda \pi^+$ 
is consistent with all theoretical predictions within $2\sigma$.

\acknowledgments
This work, based on data collected using the Belle II detector, which was built and commissioned prior to March 2019,
and data collected using the Belle detector, which was operated until June 2010,
was supported by
Higher Education and Science Committee of the Republic of Armenia Grant No.~23LCG-1C011;
Australian Research Council and Research Grants
No.~DP200101792, 
No.~DP210101900, 
No.~DP210102831, 
No.~DE220100462, 
No.~LE210100098, 
and
No.~LE230100085; 
Austrian Federal Ministry of Education, Science and Research,
Austrian Science Fund
No.~P~34529,
No.~J~4731,
No.~J~4625,
and
No.~M~3153,
and
Horizon 2020 ERC Starting Grant No.~947006 ``InterLeptons'';
Natural Sciences and Engineering Research Council of Canada, Compute Canada and CANARIE;
National Key R\&D Program of China under Contract No.~2022YFA1601903,
National Natural Science Foundation of China and Research Grants
No.~11575017,
No.~11761141009,
No.~11705209,
No.~11975076,
No.~12135005,
No.~12150004,
No.~12161141008,
No.~12475093,
and
No.~12175041,
China Postdoctoral Science Foundation GZC20240303 and 2024M760485, 
and Shandong Provincial Natural Science Foundation Project~ZR2022JQ02;
the Czech Science Foundation Grant No.~22-18469S 
and
Charles University Grant Agency project No.~246122;
European Research Council, Seventh Framework PIEF-GA-2013-622527,
Horizon 2020 ERC-Advanced Grants No.~267104 and No.~884719,
Horizon 2020 ERC-Consolidator Grant No.~819127,
Horizon 2020 Marie Sklodowska-Curie Grant Agreement No.~700525 ``NIOBE''
and
No.~101026516,
and
Horizon 2020 Marie Sklodowska-Curie RISE project JENNIFER2 Grant Agreement No.~822070 (European grants);
L'Institut National de Physique Nucl\'{e}aire et de Physique des Particules (IN2P3) du CNRS
and
L'Agence Nationale de la Recherche (ANR) under grant ANR-21-CE31-0009 (France);
BMBF, DFG, HGF, MPG, and AvH Foundation (Germany);
Department of Atomic Energy under Project Identification No.~RTI 4002,
Department of Science and Technology,
and
UPES SEED funding programs
No.~UPES/R\&D-SEED-INFRA/17052023/01 and
No.~UPES/R\&D-SOE/20062022/06 (India);
Israel Science Foundation Grant No.~2476/17,
U.S.-Israel Binational Science Foundation Grant No.~2016113, and
Israel Ministry of Science Grant No.~3-16543;
Istituto Nazionale di Fisica Nucleare and the Research Grants BELLE2,
and
the ICSC – Centro Nazionale di Ricerca in High Performance Computing, Big Data and Quantum Computing, funded by European Union – NextGenerationEU;
Japan Society for the Promotion of Science, Grant-in-Aid for Scientific Research Grants
No.~16H03968,
No.~16H03993,
No.~16H06492,
No.~16K05323,
No.~17H01133,
No.~17H05405,
No.~18K03621,
No.~18H03710,
No.~18H05226,
No.~19H00682, 
No.~20H05850,
No.~20H05858,
No.~22H00144,
No.~22K14056,
No.~22K21347,
No.~23H05433,
No.~26220706,
and
No.~26400255,
and
the Ministry of Education, Culture, Sports, Science, and Technology (MEXT) of Japan;  
National Research Foundation (NRF) of Korea Grants
No.~2016R1-D1A1B-02012900,
No.~2018R1-A6A1A-06024970,
No.~2021R1-A6A1A-03043957,
No.~2021R1-F1A-1060423,
No.~2021R1-F1A-1064008,
No.~2022R1-A2C-1003993,
No.~2022R1-A2C-1092335,
No.~RS-2023-00208693,
No.~RS-2024-00354342
and
No.~RS-2022-00197659,
Radiation Science Research Institute,
Foreign Large-Size Research Facility Application Supporting project,
the Global Science Experimental Data Hub Center, the Korea Institute of
Science and Technology Information (K24L2M1C4)
and
KREONET/GLORIAD;
Universiti Malaya RU grant, Akademi Sains Malaysia, and Ministry of Education Malaysia;
Frontiers of Science Program Contracts
No.~FOINS-296,
No.~CB-221329,
No.~CB-236394,
No.~CB-254409,
and
No.~CB-180023, and SEP-CINVESTAV Research Grant No.~237 (Mexico);
the Polish Ministry of Science and Higher Education and the National Science Center;
the Ministry of Science and Higher Education of the Russian Federation
and
the HSE University Basic Research Program, Moscow;
University of Tabuk Research Grants
No.~S-0256-1438 and No.~S-0280-1439 (Saudi Arabia), and
King Saud University,Riyadh, Researchers Supporting Project number (RSPD2024R873)  
(Saudi Arabia);
Slovenian Research Agency and Research Grants
No.~J1-9124
and
No.~P1-0135;
Ikerbasque, Basque Foundation for Science,
the State Agency for Research of the Spanish Ministry of Science and Innovation through Grant No. PID2022-136510NB-C33,
Agencia Estatal de Investigacion, Spain
Grant No.~RYC2020-029875-I
and
Generalitat Valenciana, Spain
Grant No.~CIDEGENT/2018/020;
the Swiss National Science Foundation;
The Knut and Alice Wallenberg Foundation (Sweden), Contracts No.~2021.0174 and No.~2021.0299;
National Science and Technology Council,
and
Ministry of Education (Taiwan);
Thailand Center of Excellence in Physics;
TUBITAK ULAKBIM (Turkey);
National Research Foundation of Ukraine, Project No.~2020.02/0257,
and
Ministry of Education and Science of Ukraine;
the U.S. National Science Foundation and Research Grants
No.~PHY-1913789 
and
No.~PHY-2111604, 
and the U.S. Department of Energy and Research Awards
No.~DE-AC06-76RLO1830, 
No.~DE-SC0007983, 
No.~DE-SC0009824, 
No.~DE-SC0009973, 
No.~DE-SC0010007, 
No.~DE-SC0010073, 
No.~DE-SC0010118, 
No.~DE-SC0010504, 
No.~DE-SC0011784, 
No.~DE-SC0012704, 
No.~DE-SC0019230, 
No.~DE-SC0021274, 
No.~DE-SC0021616, 
No.~DE-SC0022350, 
No.~DE-SC0023470; 
and
the Vietnam Academy of Science and Technology (VAST) under Grants
No.~NVCC.05.12/22-23
and
No.~DL0000.02/24-25.

These acknowledgements are not to be interpreted as an endorsement of any statement made
by any of our institutes, funding agencies, governments, or their representatives.

We thank the SuperKEKB team for delivering high-luminosity collisions;
the KEK cryogenics group for the efficient operation of the detector solenoid magnet and IBBelle on site;
the KEK Computer Research Center for on-site computing support; the NII for SINET6 network support;
and the raw-data centers hosted by BNL, DESY, GridKa, IN2P3, INFN, 
PNNL/EMSL, 
and the University of Victoria.

\renewcommand{\baselinestretch}{1.2}

\end{sloppypar}
\end{document}